\documentclass[10pt]{article}
\newcommand{\ghost}[1]{}
\usepackage{float}
\usepackage{amsmath}
\usepackage{graphicx}
\usepackage[table]{xcolor}
\usepackage{color}
\usepackage{orcidlink}
\usepackage{float}
\usepackage{hyperref}
\hypersetup{colorlinks=true, linkcolor=blue, citecolor=blue, urlcolor=blue}
\usepackage{caption}
\usepackage{subcaption}
\usepackage{setspace}
\usepackage{multirow}
\usepackage{longtable}
\usepackage{multicol}
\usepackage{amssymb}
\RequirePackage[numbers,sort&compress]{natbib}
\begin{document}
\baselineskip=18pt

\begin{center}
\LARGE{Astrophysical Signatures of Einstein-Skyrme Anti-de Sitter Black Holes:\\[1ex] Epicyclic Frequencies and QPO Constraints}
\end{center}

\vspace{0.2cm}

\begin{center}
{\bf Faizuddin Ahmed\orcidlink{0000-0003-2196-9622}}\footnote{\bf faizuddinahmed15@gmail.com}\\
{\it Department of Physics, The Assam Royal Global University, Guwahati, 781035, Assam, India}\\
\vspace{0.2cm}
{\bf Ahmad Al-Badawi\orcidlink{0000-0002-3127-3453}}\footnote{\bf ahmadbadawi@ahu.edu.jo}\\
{\it Department of Physics, Al-Hussein Bin Talal University, Ma'an 71111, Jordan}\\
\vspace{0.2cm}
{\bf \.{I}zzet Sakall{\i}\orcidlink{0000-0001-7827-9476}}\footnote{\bf izzet.sakalli@emu.edu.tr \,(Corresponding author)}\\
{\it Physics Department, Eastern Mediterranean University,
Famagusta 99628, North Cyprus via Mersin 10, Turkey}
\end{center}

\vspace{0.2cm}

\begin{abstract}
We study the geodesic motion and epicyclic oscillations of massive test particles around a static, spherically symmetric black hole (BH) solution of the Einstein--Skyrme (ES) theory in Anti-de Sitter (AdS) spacetime. The lapse function of this BH depends on the Skyrme coupling $\eta$, a charge-like parameter $Q$ inherited from the Skyrme term, and the cosmological constant $\Lambda<0$. We first map out the horizon structure and identify three regimes-non-extremal BH (NEBH), extremal BH (EBH), and naked BH (NBH)-showing that the NEBH $\to$ EBH $\to$ NBH transition is governed by $Q$ rather than $\eta$, which enters $f(r)$ only as a constant shift. We then derive the effective potential (EP), locate the innermost stable circular orbit (ISCO), and compute the radiative efficiency, finding that $\mathcal{E}_{\rm ISCO}>1$ in AdS renders the standard Novikov-Thorne formula negative. The corrected radial epicyclic frequency $\Omega_r$ reveals a distinctive AdS signature: $\nu_r$ grows at large $r$ and overtakes the orbital frequency $\nu_\phi$, causing the periastron precession frequency $\nu_p = \nu_\varphi - \nu_r$ to change sign-a feature absent in asymptotically flat geometries. Adopting the relativistic precession (RP) model for quasi-periodic oscillations (QPOs), we perform a Markov chain Monte Carlo (MCMC) analysis using twin-peak QPO data from XTE~J1550-564, GRO~J1655-40, Sgr~A$^*$, and M82~X-1. The posteriors converge to $Q\approx 0.6$ across all sources, with orbital radii near $r\approx 4.2\,M$ and masses consistent with independent estimates, demonstrating that the ES-AdS BH accommodates the observed frequency pairs within physically motivated parameter ranges.
\end{abstract}
 
\noindent\textit{Keywords:} Einstein-Skyrme black holes; Anti-de Sitter spacetime; epicyclic frequencies; quasi-periodic oscillations; MCMC

\section{Introduction}\label{isec1}
 
The direct imaging of the supermassive BH shadows in M87$^*$ \cite{EHTL1,EHTL6} and Sgr~A$^*$ \cite{EHTL12,EHTL17} by the Event Horizon Telescope (EHT) collaboration has opened a new window for testing general relativity (GR) in the strong-field regime.\ghost{} These observations, combined with gravitational wave (GW) detections by the LIGO-Virgo-KAGRA network \cite{LIGOVirgo2016,LIGOVirgo2021}, place increasingly tight constraints on the near-horizon geometry of compact objects. At the same time, X-ray timing missions such as RXTE, NICER, and the forthcoming eXTP continue to refine measurements of QPOs in accreting BH systems \cite{Klis2000,Remillard2006}, providing frequency data that probe the orbital and epicyclic structure of the underlying spacetime.
 
Among the theoretical frameworks that extend GR, the Skyrme model occupies a distinctive position.\ghost{} Originally introduced by Skyrme \cite{Skyrme1961,Skyrme1962} as a nonlinear sigma model of pions in which baryons arise as topological solitons, it was subsequently placed on firm theoretical footing through its connection to large-$N_c$ QCD \cite{Witten1983,Adkins1983}. Coupling the Skyrme field to gravity yields self-gravitating skyrmion configurations \cite{Luckock1986,Droz1991,Kleihaus1995} that have been explored in a variety of contexts, from neutron star interiors \cite{Nelmes2012,Adam2015} to BH hair \cite{Gudnason2017,Canfora2019,Brihaye2006}.\ghost{} Of particular interest is the family of exact, static, and spherically symmetric BH solutions found by Canfora and collaborators \cite{Canfora2013,Canfora2014,Canfora2018} in the ES theory with a negative cosmological constant. In these solutions the Skyrme field, adopted with a hedgehog ansatz on $S^3$, contributes a $Q^2/r^2$ term to the lapse function whose radial dependence mirrors the Maxwell term in the Reissner-Nordstr\"om (RN) metric, yet whose coefficient is fixed by the pion decay constant $F_{\pi}$ and the Skyrme coupling $e$ rather than being a free integration constant.
 
AdS spacetimes have attracted sustained attention for reasons that go beyond the celebrated AdS/CFT correspondence \cite{Maldacena1998,Witten1998AdS}.\ghost{} In BH physics, the negative cosmological constant introduces a confining potential wall at large $r$ that qualitatively modifies the causal structure, thermodynamic phase transitions \cite{Kubiznak2012,Gunasekaran2012,Hawking1983}, and geodesic dynamics compared to the asymptotically flat case \cite{Stuchlik2000,Cruz2005}. For instance, circular orbits in AdS--Schwarzschild backgrounds exist only within a bounded radial range, and the specific energy on such orbits grows without bound-properties that bear directly on the interpretation of accretion disk observables.\ghost{}
 
QPOs detected in the X-ray light curves of low-mass X-ray binaries (LMXBs) and active galactic nuclei (AGN) manifest as sharp peaks in the power spectrum at frequencies ranging from millihertz (for supermassive BHs) to hundreds of hertz (for stellar-mass systems) \cite{Klis2000,Remillard2006}.\ghost{} The twin-peak, or high-frequency (HF), QPOs are of special interest because their frequency ratio clusters near small integers, suggesting a resonant or precession-based origin tied to the strong gravitational field near the ISCO \cite{Abramowicz2001,Torok2005}. Several competing models exist for identifying the upper and lower QPO frequencies $(\nu_U,\nu_L)$ with combinations of the orbital, radial, and vertical epicyclic frequencies \cite{Stella1998,Stella1999,Rezzolla2003}.\ghost{} The RP model, which sets $\nu_U = \nu_\varphi$ and $\nu_L = \nu_\varphi - \nu_r$, has proven particularly successful for Schwarzschild- and Kerr-like BHs and has been extended to a range of modified gravity scenarios \cite{Bambi2012,Rayimbaev2021,Boshkayev2023, Boos2026, Kolos2020, Shahzadi2021, Mustafa2025, Ashraf2025a,Ashraf2025b}. Alternative multi-resonance approaches have also been developed \cite{StuchlikKonar2008,StuchlikKotrlova2012}. Significant contributions to the understanding of QPO phenomena have been made by Stuchl{\'\i}k et al. \cite{Stuchlik2013, Kotrlova2014, Stuchlik2016, Stuchlik2021}.
 
Recent studies have applied the QPO framework to BHs embedded in dark matter halos \cite{Shaymatov2021,StuchlikVrba2021,Davlataliev2024,AhmedAlBadawiSakalli2025f}, regular BH models \cite{Toshmatov2019,Stuchlik2021,AhmedAlBadawiSakalli2026a}, BHs with nonlinear electrodynamics \cite{Alloqulov2023,AhmedAlBadawiSilva2026,AhmedFathiAlBadawi2026,AydinerSucuSakalli2025}, and quantum-corrected or hairy BH configurations \cite{Franchini2017,AlBadawiAhmedDonmez2025}.\ghost{} In each case, the deformation parameters of the metric leave measurable imprints on the epicyclic frequencies and can be constrained through MCMC or $\chi^2$ fitting to observed QPO data \cite{Ghorani2024}. This approach has been gainfully applied to sources spanning the stellar-mass \cite{Remillard2002,Strohmayer2001}, intermediate-mass \cite{Pasham2014}, and supermassive \cite{Genzel2003} BH regimes. In a closely related line of work, AdS BH solutions coupled to exotic matter sources have been studied in \cite{AhmedAlBadawiSakalli2025h,AhmedAlBadawiSakalli2026b,AhmedAlBadawiSakalli2025g}, where the interplay between the cosmological constant and additional field content produces observable modifications to the shadow, QNM spectrum, and thermodynamic phase structure. Possible signature of the magnetic fields related to quasi-periodic oscillations observed in microquasars has been reported in \cite{Kolos2017}. 

Despite this progress, the orbital dynamics and QPO phenomenology of ES-AdS BHs have not been explored.\ghost{} The present work aims to fill this gap. We carry out a detailed study of the geodesic structure, horizon taxonomy, EP profiles, ISCO properties, and epicyclic frequencies of the ES-AdS BH, and we confront the theoretical predictions with observational QPO data through a Bayesian MCMC analysis. A noteworthy outcome of our investigation is the identification of a distinctive AdS signature in the periastron precession frequency: $\nu_p = \nu_\varphi - \nu_r$ changes sign at a finite radius because the confining AdS potential causes $\nu_r$ to grow and eventually exceed $\nu_\varphi$-a behavior with no counterpart in asymptotically flat spacetimes.\ghost{}
 
The paper is organized as follows. In Sec.~\ref{isec2} we present the ES-AdS BH solution, define the relevant parameters, and analyze the horizon structure with a classification into NEBH, EBH, and NBH configurations. Section~\ref{isec3} is devoted to the dynamics of massive test particles: we derive the EP, determine the specific energy and angular momentum on circular orbits, locate the ISCO, and compute the radiative efficiency.\ghost{} In Sec.~\ref{isec4} we obtain the epicyclic frequencies, examine the periastron precession, and perform the MCMC analysis of QPO data from four astrophysical sources. We summarize our findings and discuss future directions in Sec.~\ref{isec5}.

{\color{black}
\section{Einstein-Skyrme AdS Black Holes}\label{isec2}
 
The Skyrme model, originally proposed by Skyrme \cite{Skyrme1961,Skyrme1962} as a nonlinear field theory of pions, has proven to be a fruitful framework for describing baryons as topological solitons.\ghost{} Coupling this model to GR produces gravitating skyrmion solutions, among which static and spherically symmetric BH configurations have attracted particular attention \cite{Luckock1986,Droz1991,Kleihaus1995}. When the Skyrme field is adopted with a hedgehog ansatz on $S^3$, the coupled ES field equations admit analytic BH solutions in AdS spacetime, as shown by Canfora and collaborators \cite{Canfora2013,Canfora2014,Canfora2018}.\ghost{} These solutions are of special interest because the Skyrme term modifies the near-horizon geometry in a way that resembles, yet is physically distinct from, the electromagnetic contribution in the RN solution.
 
The static, spherically symmetric ES-AdS BH line element reads \cite{Canfora2013,Canfora2014,Canfora2018}
\begin{equation}
    ds^2 = -f(r)\,dt^2 + \frac{dr^2}{f(r)} + r^2 \left(d\theta ^2 + \sin ^2{\theta }\,d\varphi ^2\right),\label{metric}
\end{equation}
where the lapse function takes the form
\begin{equation}
     f(r)= 1 - \eta^2 - \frac{r_s}{r} + \frac{Q^2}{r^2} - \frac{\Lambda}{3}\, r^2,\label{f-of-r}
\end{equation}
with the identifications\ghost{}
\begin{equation}
    \eta^2 \equiv 8\pi G K,\quad r_s \equiv 2 G M,\quad Q^2 \equiv 4\pi G K \lambda.\label{param-def}
\end{equation}
Here $M$ is the BH mass, $G$ denotes Newton's constant, $\Lambda < 0$ is the cosmological constant (encoding the AdS curvature radius via $\Lambda = -3/\ell^2$), while the Skyrme sector is characterized by
\begin{equation}
    K = \frac{F_\pi^2}{4}, \qquad \lambda = \frac{4}{e^2 F_\pi^2},\label{skyrme-couplings}
\end{equation}
where the pion decay constant $F_\pi$ and the dimensionless coupling $e$ are fixed phenomenologically as \cite{Canfora2014,Adkins1983}\ghost{}
\begin{equation}
    F_\pi = 0.141~\text{GeV}, \qquad 5 \le e \le 7.\label{pheno-values}
\end{equation}
 
Some physical features of the metric \eqref{f-of-r} deserve attention. The constant deficit $(1-\eta^2)$ plays the role of a solid angle deficit, much like the global monopole spacetime of Barriola and Vilenkin \cite{Barriola1989}; setting $Q = 0$ and $\Lambda = 0$ reduces \eqref{f-of-r} to precisely that case.\ghost{} The $Q^2/r^2$ contribution arising from the Skyrme term shares the same radial dependence as the Maxwell term in the RN metric; however, $Q^2$ here is not an independent integration constant but is fixed by the coupling parameters of the theory through Eq.~\eqref{skyrme-couplings}. This distinction has direct consequences for the BH thermodynamics and the parameter space of the solution. The AdS term $r^2/\ell^2$ (with $\Lambda=-3/\ell^2$) ensures that $f(r) \to +\infty$ as $r \to \infty$, which guarantees the existence of an outer EH for sufficiently large $M$.

An important subtlety concerns the interrelation between $\eta$ and $Q$. Substituting Eq.~\eqref{skyrme-couplings} into the definitions \eqref{param-def} yields\ghost{}
\begin{equation}
    \eta^2 = 2\pi G F_\pi^2, \qquad Q^2 = \frac{4\pi G}{e^2}.\label{eta-Q-relation}
\end{equation}
The pion decay constant $F_\pi$ drops out entirely from $Q^2$, so the two metric parameters $\eta$ and $Q$ are governed by \emph{independent} physical couplings: $\eta$ tracks $F_\pi$ (the sigma-model kinetic term), while $Q$ tracks the Skyrme coupling $e$ (the quartic stabilizing term).\ghost{} The limit $\eta \to 0$ with $Q \neq 0$ is therefore physically meaningful---it corresponds to the BPS Skyrme regime \cite{Adam2015} in which the sigma-model term is suppressed ($F_\pi \to 0$) while the quartic Skyrme term survives at finite $e$. The opposite corner, $Q \to 0$ with $\eta \neq 0$, requires $e \to \infty$ and yields a global-monopole--AdS spacetime with no Skyrme charge.\ghost{} Only when both $F_\pi = 0$ and $e = \infty$ (i.e.\ $\eta = Q = 0$) does the full Skyrme sector decouple, recovering the Schwarzschild-AdS solution. Throughout the rest of this paper we treat $\eta$ and $Q$ as independent parameters in the metric \eqref{f-of-r}, keeping in mind that specific numerical values can always be mapped back to $(F_\pi, e)$ via Eq.~\eqref{eta-Q-relation}.

\subsection{Horizon structure}\label{isec2a}
 
The EH locations are obtained from $f(r_h)=0$, which, after multiplying by $r_h^2$, yields the quartic equation\ghost{}
\begin{equation}
    \frac{r_h^4}{\ell^2} + (1-\eta^2)\,r_h^2 - r_s\,r_h + Q^2 = 0.\label{quartic-horizon}
\end{equation}
For $\Lambda = 0$, the equation reduces to a quadratic in $r_h$ with the familiar discriminant $\Delta = r_s^2 - 4\,(1-\eta^2)\,Q^2$, giving two, one, or zero positive roots depending on the sign of $\Delta$. For $\Lambda < 0$ the quartic nature of \eqref{quartic-horizon} enriches the causal structure: generically, the solution admits a Cauchy horizon $r_-$ and an EH $r_+$, which we denote the non-extremal BH (NEBH) case; a single degenerate horizon where $f(r_h)=f'(r_h)=0$, the extremal BH (EBH) case; or no horizon at all, the naked BH (NBH) case.\ghost{}
 
To map out the parameter space, we solve $f(r_h)=0$ numerically via a sign-change root-finding algorithm across a fine radial grid. The resulting classification is collected in Table~\ref{tab:horizons}. Several trends are worth noting. Increasing $\eta$ shifts the outer horizon outward, since the deficit $(1-\eta^2)$ weakens the effective gravitational pull. Increasing $Q$ pushes the Cauchy horizon outward and can merge the two horizons into an EBH or eliminate them entirely (NBH). A more negative $\Lambda$ (stronger AdS curvature) enlarges the outer horizon through the $r^2/\ell^2$ term.\ghost{}
 
A noteworthy feature of \eqref{f-of-r} is that $\eta$ enters only through the constant shift $(1-\eta^2)$, so it drops out of $f'(r)$. The location where $f$ attains its minimum between the horizons is therefore determined solely by $Q$ and $\Lambda$ (and $M$), while $\eta$ translates the entire curve vertically. Because both the $Q^2/r^2$ repulsion at small $r$ and the AdS potential $|\Lambda|r^2/3$ at large $r$ drive $f\to+\infty$, any local minimum that lies below zero always generates two zero crossings.\ghost{} As a consequence, the NEBH $\to$ EBH $\to$ NBH transition is triggered by increasing $Q$ past a critical value $Q_{\rm ext}$, where the repulsive $1/r^2$ term lifts the minimum above zero. The extremal condition $f(r_h) = f'(r_h) = 0$ can be reduced analytically to
\begin{equation}
    M = r_h\!\left(1-\eta^2 - \frac{2\Lambda}{3}\,r_h^2\right),\qquad Q_{\rm ext}^2 = r_h\!\left(M - \frac{\Lambda}{3}\,r_h^3\right),\label{ext-cond}
\end{equation}
which yields both $r_{\rm ext}$ and $Q_{\rm ext}$ once $\eta$ and $\Lambda$ are specified.\ghost{}

\setlength{\tabcolsep}{12pt}
\renewcommand{\arraystretch}{1.6}
\begin{longtable}{|c|c|c|c|c|}
\hline
\rowcolor{orange!50}
\textbf{$\eta$} & \textbf{$Q$} & \textbf{$\Lambda$} & \textbf{Horizon(s) $r_h$} & \textbf{Configuration} \\
\hline
\endfirsthead
\hline
\rowcolor{orange!50}
\textbf{$\eta$} & \textbf{$Q$} & \textbf{$\Lambda$} & \textbf{Horizon(s) $r_h$} & \textbf{Configuration} \\
\hline
\endhead
0.00 & 0.0000 & 0.00 & $[2.0000]$ & Single Horizon BH \\
\hline
0.00 & 0.5000 & 0.00 & $[0.1340,\ 1.8660]$ & NEBH \\
\hline
0.00 & 1.0000 & 0.00 & $[\,]$ & NBH \\
\hline
0.20 & 0.0000 & 0.00 & $[2.0833]$ & Single Horizon BH \\
\hline
0.20 & 0.7000 & 0.00 & $[0.2836,\ 1.7997]$ & NEBH \\
\hline
0.50 & 1.0000 & 0.00 & $[0.6667,\ 2.0000]$ & NEBH \\
\hline
0.00 & 0.0000 & $-0.03$ & $[1.9283]$ & Single Horizon BH \\
\hline
0.00 & 0.7000 & $-0.03$ & $[0.2859,\ 1.6590]$ & NEBH \\
\hline
0.00 & 1.0000 & $-0.03$ & $[\,]$ & NBH \\
\hline
0.10 & 0.3000 & $-0.03$ & $[0.0460,\ 1.9028]$ & NEBH \\
\hline
0.10 & 0.7000 & $-0.03$ & $[0.2853,\ 1.6775]$ & NEBH \\
\hline
0.10 & 1.0000 & $-0.03$ & $[0.9809,\ 1.0000]$ & NEBH \\
\hline
0.30 & 0.5000 & $-0.03$ & $[0.1331,\ 1.9741]$ & NEBH \\
\hline
0.50 & 1.0000 & $-0.03$ & $[0.6687,\ 1.8653]$ & NEBH \\
\hline
0.00 & 0.0000 & $-0.10$ & $[1.8042]$ & Single Horizon BH \\
\hline
0.10 & 0.7000 & $-0.10$ & $[0.2854,\ 1.5744]$ & NEBH \\
\hline
0.10 & 1.0000 & $-0.10$ & $[\,]$ & NBH \\
\hline
0.50 & 1.0000 & $-0.10$ & $[0.6736,\ 1.6602]$ & NEBH \\
\hline
\caption{Representative horizon configurations of the ES-AdS BH for $M=1$. Single Horizon BH: one root with $f'(r_h) \neq 0$ (Schwarzschild-like); NEBH: two distinct horizons (Cauchy and event); NBH: no horizon (naked singularity). The full parameter scan covers $\eta \in \{0,0.1,0.2,0.3,0.5\}$, $Q \in \{0,0.3,0.5,0.7,1.0\}$, and $\Lambda \in \{0,-0.03,-0.10\}$.}\label{tab:horizons}
\end{longtable}

\begin{figure}[ht!]
    \centering
    \includegraphics[width=0.67\linewidth]{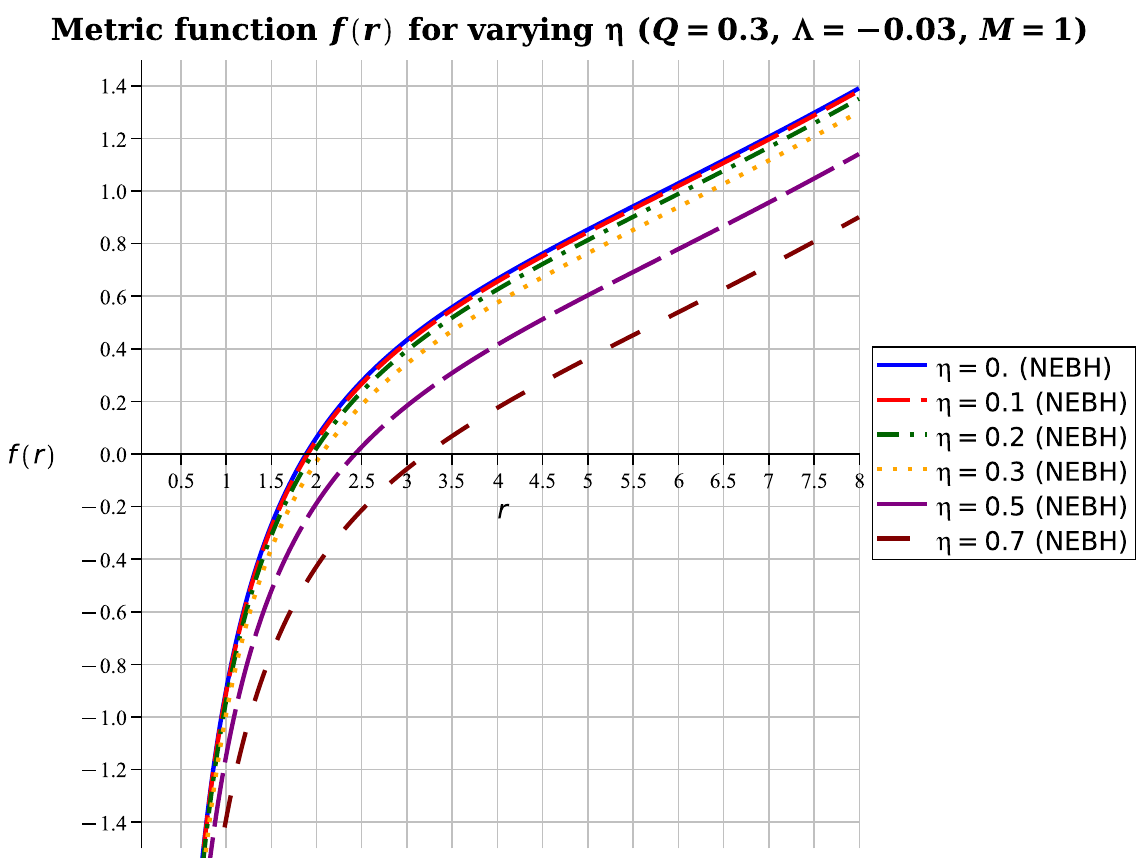}
    \caption{Metric function $f(r)$ for several values of the Skyrme coupling parameter $\eta$, with $Q = 0.3$, $\Lambda = -0.03$, and $M = 1$. All curves remain in the NEBH regime: the outer EH moves outward and the inter-horizon gap widens with increasing $\eta$, while two horizons persist due to the combined $Q^2/r^2$ and AdS barriers.}\label{fig:f_vs_eta}
\end{figure}
 
Representative profiles of $f(r)$ are displayed in Figs.~\ref{fig:f_vs_Q}-\ref{fig:f_configs}.\ghost{} Figure~\ref{fig:f_vs_Q} is the main figure of this section and shows the behavior obtained by varying $Q$ at fixed $\eta = 0.1$ and $\Lambda = -0.03$: small $Q$ values produce well-separated Cauchy and event horizons (NEBH), the critical value $Q = Q_{\rm ext}$ yields a degenerate horizon tangent to the axis (EBH), and $Q = 1.5$ lies entirely above zero near the origin, with no zero crossing (NBH). In Fig.~\ref{fig:f_vs_eta}, we fix $Q = 0.3$ and $\Lambda = -0.03$ while varying $\eta$. All curves remain in the NEBH regime: the outer EH shifts outward and the inter-horizon gap widens as $\eta$ grows, but two distinct horizons survive for every $\eta$ tested, consistent with the argument given above.\ghost{} The influence of $\Lambda$ is illustrated in Fig.~\ref{fig:f_vs_Lambda}; the AdS potential well deepens with $|\Lambda|$, pulling the outer horizon to larger radii and broadening the allowed region for stable circular orbits. Figure~\ref{fig:f_configs} collects four characteristic profiles-Schwarzschild-AdS (NEBH), a generic NEBH, EBH at $Q = Q_{\rm ext}$, and NBH-in a single panel to make the full progression explicit.
 
\begin{figure}[ht!]
    \centering
    \includegraphics[width=0.67\linewidth]{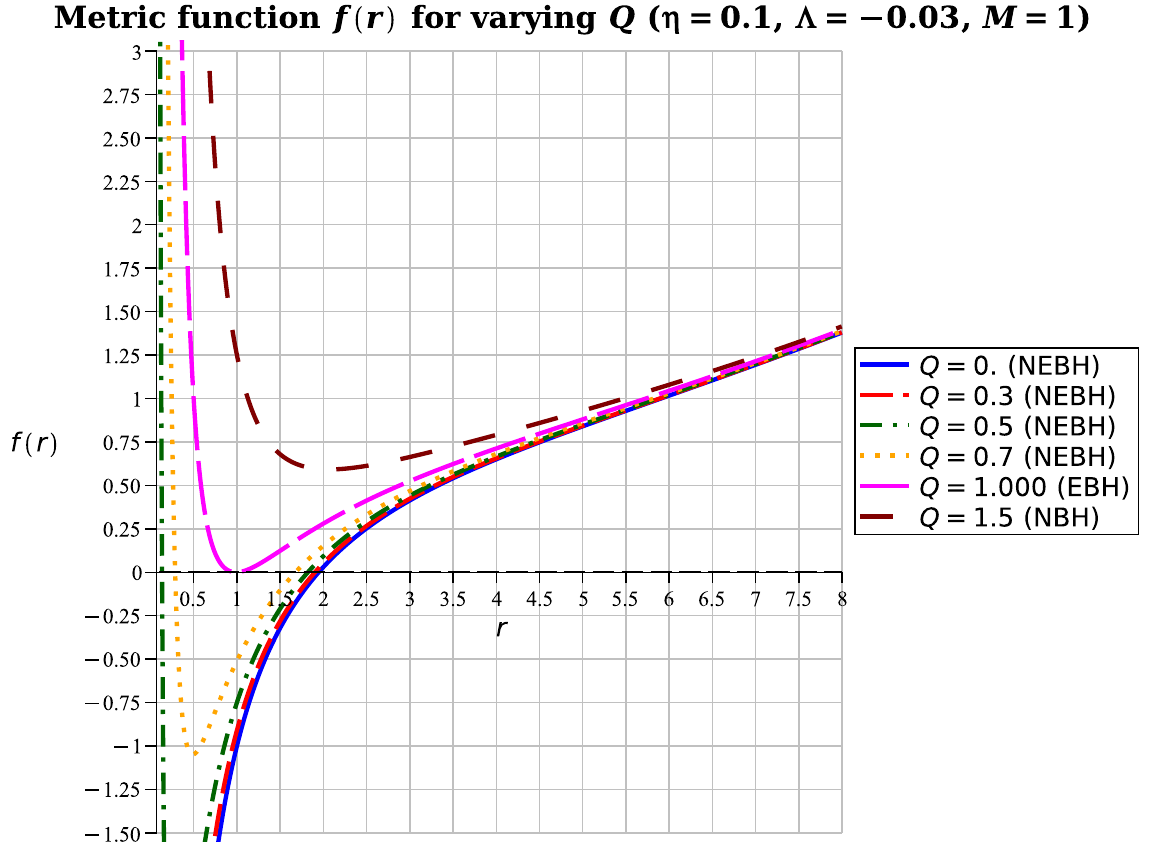}
    \caption{Metric function $f(r)$ for varying Skyrme charge parameter $Q$, with $\eta = 0.1$, $\Lambda = -0.03$, and $M = 1$. NEBH curves ($Q < Q_{\rm ext}$) exhibit two zero crossings (Cauchy and event horizons); the EBH curve ($Q = Q_{\rm ext}$) is tangent to $f=0$; the NBH curve ($Q = 1.5$) shows no zero crossing, indicating the absence of any horizon.\ghost{}}\label{fig:f_vs_Q}
\end{figure}
 
\begin{figure}[ht!]
    \centering
    \includegraphics[width=0.7\linewidth]{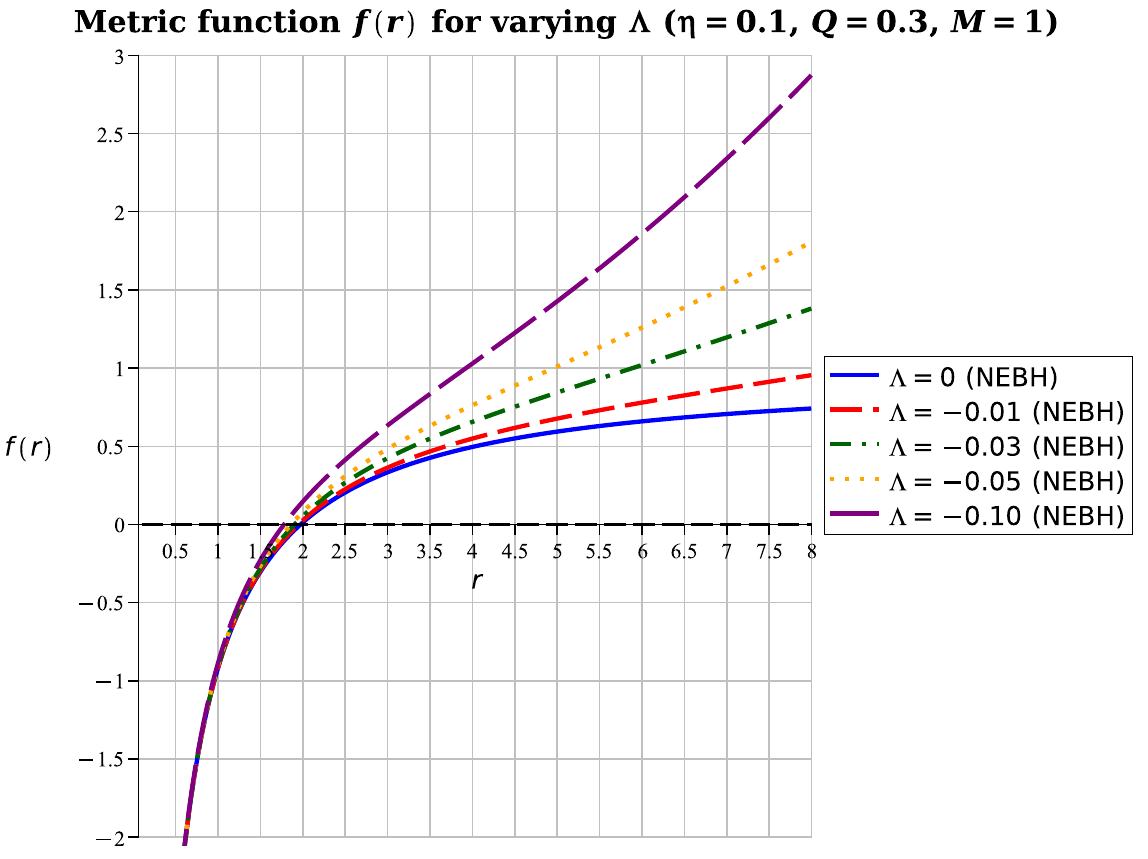}
    \caption{Metric function $f(r)$ for different values of the cosmological constant $\Lambda$, with $\eta = 0.1$, $Q = 0.3$, and $M = 1$. All configurations shown are NEBH. More negative $\Lambda$ strengthens the AdS potential well, pushing the outer EH to larger radii.\ghost{}}\label{fig:f_vs_Lambda}
\end{figure}
 
\begin{figure}[ht!]
    \centering
    \includegraphics[width=0.7\linewidth]{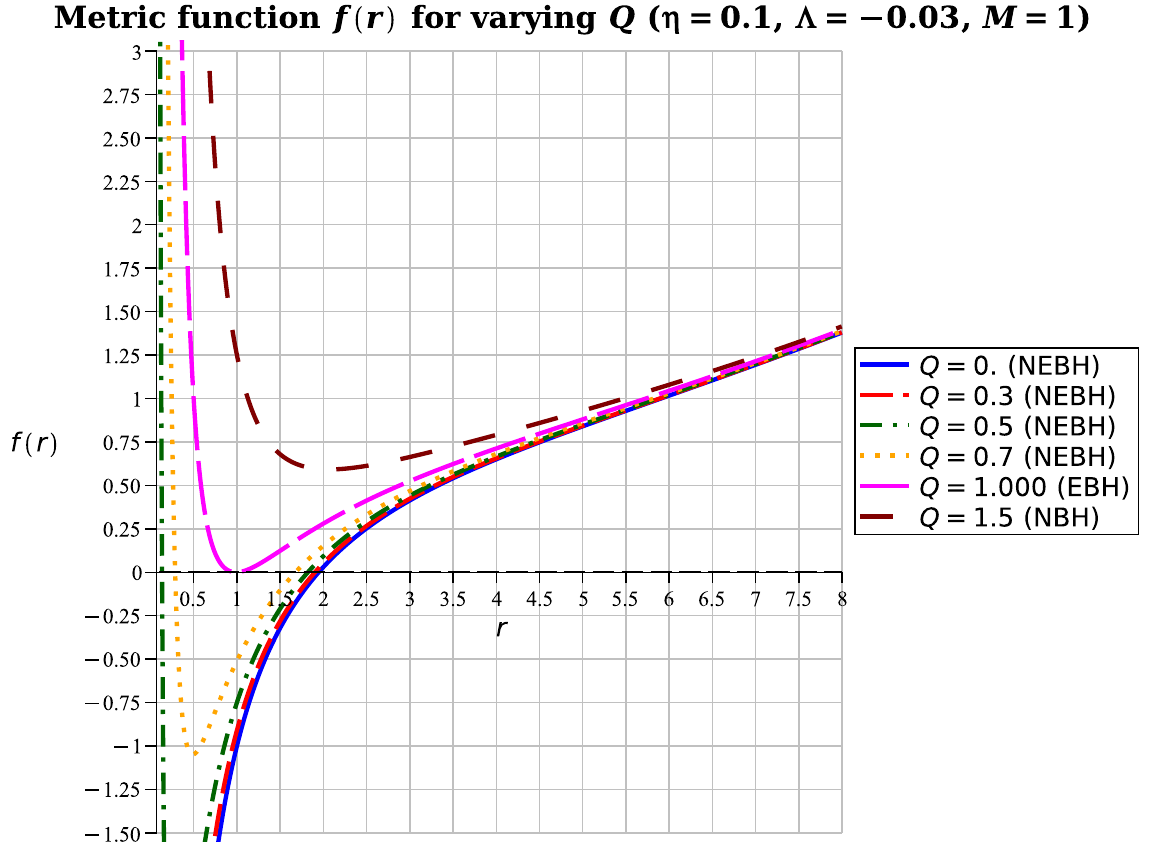}
    \caption{Characteristic configurations of the ES-AdS BH with $\Lambda = -0.03$ and $M = 1$: Schwarzschild-AdS (NEBH, $\eta = 0$, $Q = 0$), generic NEBH ($\eta = 0.1$, $Q = 0.3$), EBH ($\eta = 0.1$, $Q = Q_{\rm ext}$), and NBH ($\eta = 0.1$, $Q = 1.5$).}\label{fig:f_configs}
\end{figure}

\section{Dynamics of Test Particles}\label{isec3}

In this section we study the motion of massive neutral test particles in the ES-AdS BH background \eqref{metric}-\eqref{f-of-r}.\ghost{} The geodesic structure of the spacetime encodes information about the gravitational field that is directly linked to observational signatures-orbital frequencies, accretion disk properties, and the location of the ISCO, all of which can be constrained by EHT data and X-ray timing observations \cite{Chandrasekhar1984,RMW1984}.

\subsection{Equations of motion}\label{isec3a}

The Lagrangian for a test particle of mass $m$ moving along a time-like geodesic with affine parameter $\tau$ reads \cite{Chandrasekhar1984}\ghost{}
\begin{equation}
    \mathbb{L}=\frac{1}{2}\, m\, g_{\mu\nu}\, \frac{dx^{\mu}}{d\tau}\, \frac{dx^{\nu}}{d\tau},\label{lagrangian}
\end{equation}
where $g_{\mu\nu}$ is the metric tensor for the (1+3)-dimensional black hole space-time with $\mu, \nu=0,\cdot, 3$.

Substituting the metric \eqref{metric} and restricting to the equatorial plane $\theta = \pi/2$ (which is consistent since the spacetime is spherically symmetric), the two Killing symmetries $\partial_t$ and $\partial_\varphi$ yield the conserved specific energy and specific angular momentum\ghost{}
\begin{equation}
    \mathcal{E} = f(r)\,\frac{dt}{d\tau}, \qquad \mathcal{L} = r^2\,\frac{d\varphi}{d\tau}.\label{conserved}
\end{equation}
Here $\mathcal{E}$ and $\mathcal{L}$, respectively, are the conserved energy and angular momentum per unit mass of the test particles.

Massive test particles obeys the normalization condition $g_{\mu\nu}\,u^\mu u^\nu = -1$, where $u^\mu=\mathrm{d}x^{\mu } / \mathrm{d}\tau=\{u^{t},u^r,0,u^{\phi }\}$ is the four-velocity of the particle. This implies that for the equatorial motion, the radial component can be written as
\begin{equation}
    \left(\frac{dr}{d\tau}\right)^2 + U_{\rm eff}(r) = \mathcal{E}^2,\label{radial-eom}
\end{equation}
which is equivalent to the one-dimensional equation of motion of a unit mass particle and $U_{\rm eff}(r)$ is the effective potential (EP) takes the following form
\begin{equation}
    U_{\rm eff}(r) = \left(1+\frac{\mathcal{L}^2}{r^2}\right) f(r).\label{Veff}
\end{equation}
\ghost{}The interplay between the centrifugal barrier $\mathcal{L}^2/r^2$, the Skyrme charge term $Q^2/r^2$ inside $f(r)$, and the AdS potential $|\Lambda|r^2/3$ produces a rich landscape of bound and unbound orbits, which we now explore.

\subsection{Effective potential profiles}\label{isec3b}

Figure~\ref{fig:ep_vs_L} displays $U_{\rm eff}(r)$ for several values of $\mathcal{L}$ at fixed $\eta=0.1$, $Q=0.3$, and $\Lambda=-0.03$. For low angular momentum the potential is nearly monotonic and the particle falls into the BH; as $\mathcal{L}$ grows, a local maximum and a local minimum develop, separated by a potential well that supports stable circular orbits.\ghost{} The minimum of $U_{\rm eff}$ corresponds to the stable circular orbit at that $\mathcal{L}$, and its value gives $\mathcal{E}^2$ on the orbit.  A distinctive feature of the AdS background is the rise of $U_{\rm eff}$ at large $r$, which acts as a confining wall and prevents particles from escaping to infinity.

The dependence on the Skyrme coupling $\eta$ is shown in Fig.~\ref{fig:ep_vs_eta}: larger $\eta$ lowers the peak of the potential barrier and shifts the minimum outward, implying that the BH captures particles from a wider range of impact parameters.\ghost{} The effect of $Q$ is exhibited in Fig.~\ref{fig:ep_vs_Q}; increasing $Q$ deepens the inner well and moves the potential minimum to smaller radii, reflecting the strengthened short-range repulsion from the $Q^2/r^2$ term.
\begin{figure}[ht!]
    \centering
    \includegraphics[width=0.7\linewidth]{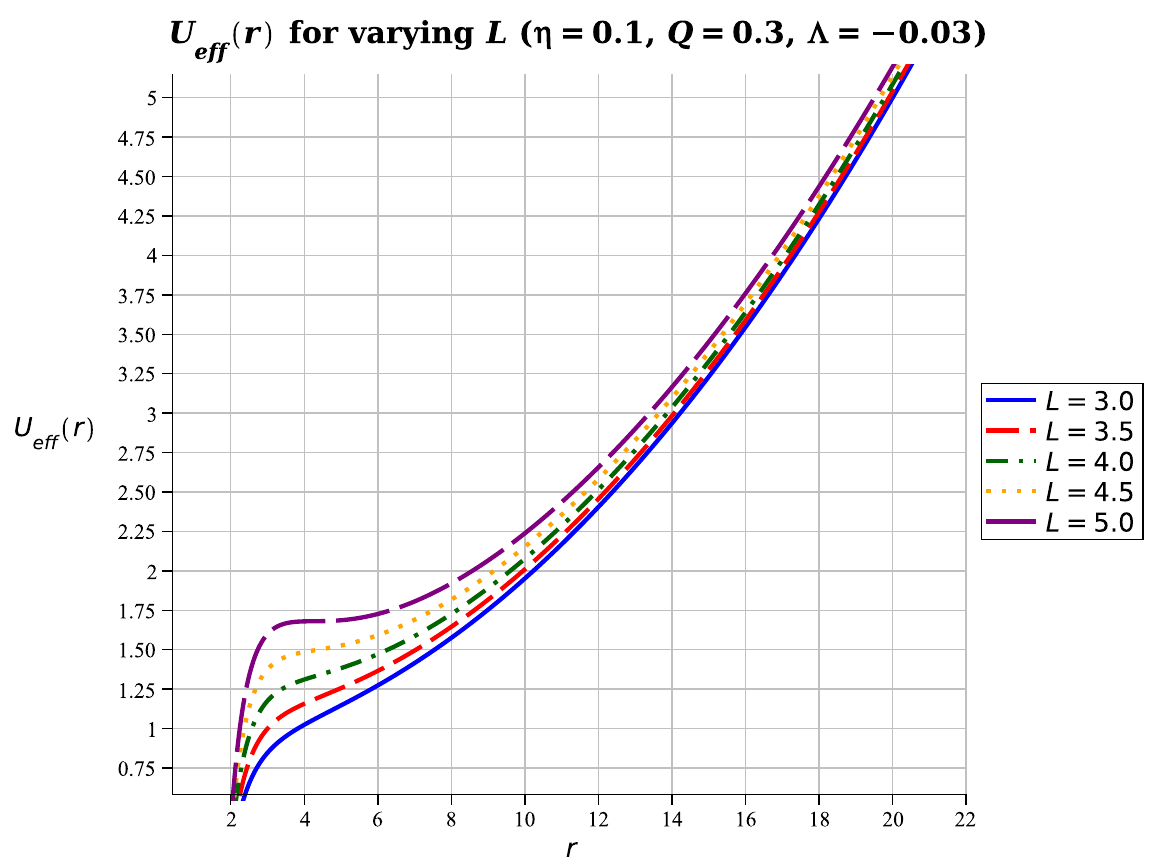}
    \caption{EP $U_{\rm eff}(r)$ for varying $\mathcal{L}$, with $\eta = 0.1$, $Q = 0.3$, $\Lambda = -0.03$, and $M = 1$. Higher angular momentum widens the potential well and supports stable circular orbits. The rise at large $r$ is the AdS confining wall.}\label{fig:ep_vs_L}
\end{figure}

\begin{figure}[ht!]
    \centering
    \includegraphics[width=0.7\linewidth]{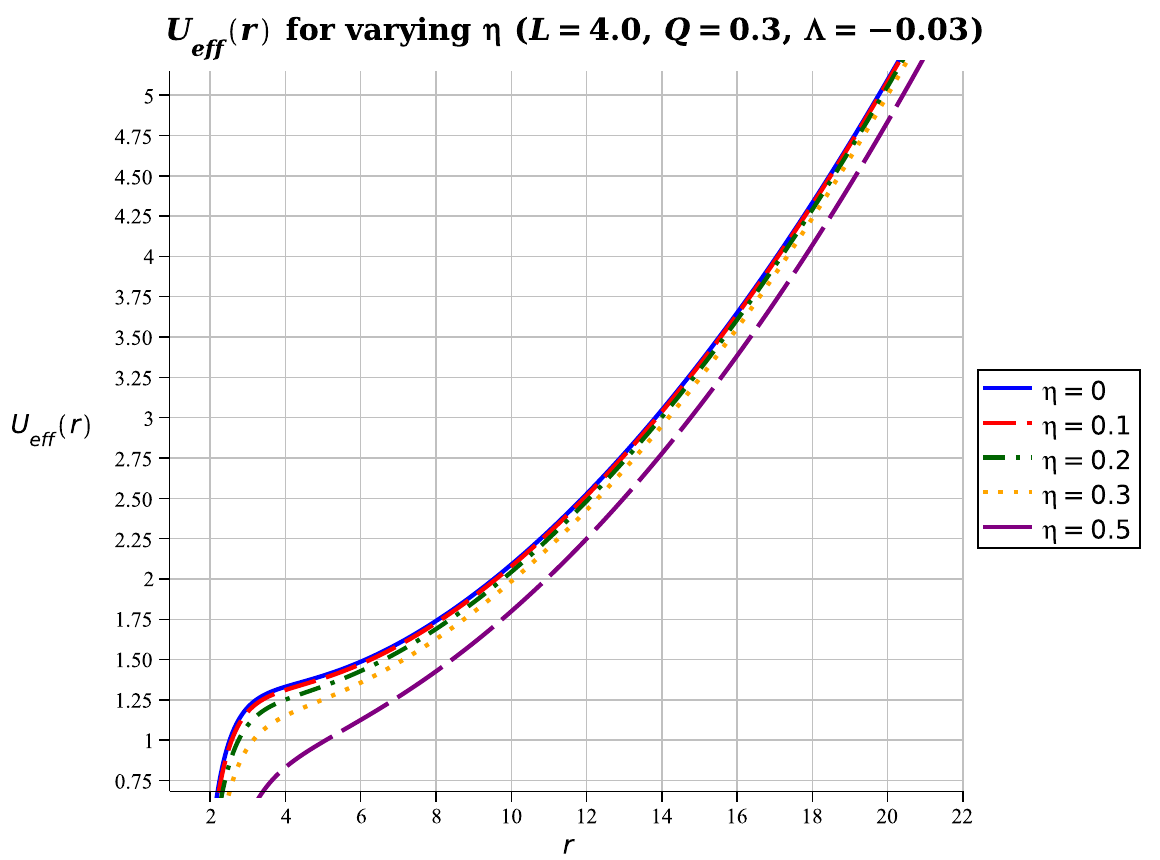}
    \caption{EP $U_{\rm eff}(r)$ for varying $\eta$, with $\mathcal{L} = 4$, $Q = 0.3$, $\Lambda = -0.03$, and $M = 1$. Larger $\eta$ lowers the potential barrier and pushes the minimum outward.\ghost{}}\label{fig:ep_vs_eta}
\end{figure}

\begin{figure}[ht!]
    \centering
    \includegraphics[width=0.7\linewidth]{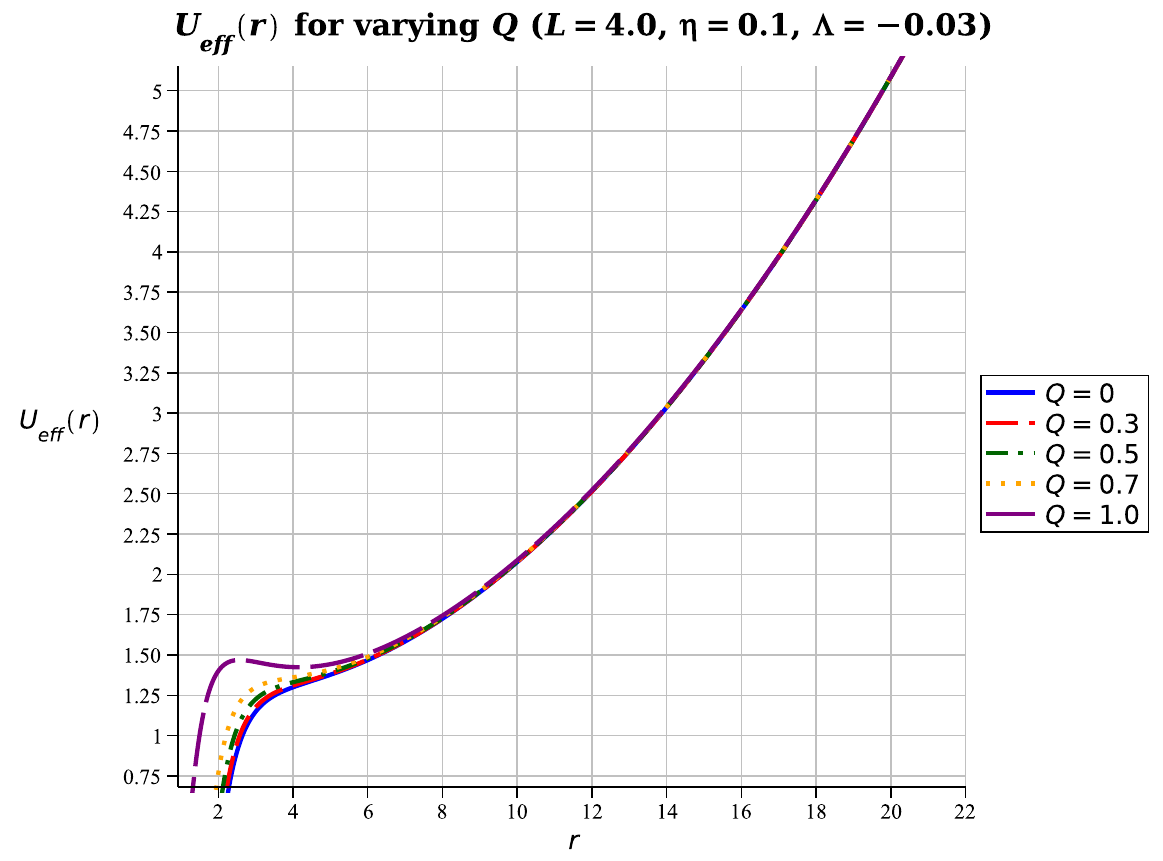}
    \caption{EP $U_{\rm eff}(r)$ for varying $Q$, with $\mathcal{L} = 4$, $\eta = 0.1$, $\Lambda = -0.03$, and $M = 1$. The $Q^2/r^2$ repulsion deepens the inner well and shifts the minimum to smaller $r$.}\label{fig:ep_vs_Q}
\end{figure}

\subsection{Effective radial force}\label{isec3b2}

The effective radial force experienced by a massive test particle in the ES-AdS BH gravitational field equals the negative gradient of the EP.\ghost{} It is defined by
\begin{equation}
    \mathcal{F} = -\frac{1}{2}\,\frac{\partial U_{\rm eff}}{\partial r}.\label{force-def}
\end{equation}
Substituting the EP \eqref{Veff} and carrying out the differentiation yields
\begin{equation}
    \mathcal{F} = -\frac{1}{2}\left(\frac{r_s}{r^2} - \frac{2Q^2}{r^3} - \frac{2\Lambda}{3}\,r\right) + \frac{\mathcal{L}^2}{r^3}\left(1 - \eta^2 - \frac{3\,r_s}{2\,r} + \frac{2Q^2}{r^2}\right).\label{force-expr}
\end{equation}
The first group of terms represents the radial force on a particle with zero angular momentum: the $r_s/r^2$ piece is the usual Newtonian attraction, the $Q^2/r^3$ repulsion originates in the Skyrme charge, and the $\Lambda r$ term reflects the AdS confining potential.\ghost{} The second group, proportional to $\mathcal{L}^2$, encodes the centrifugal corrections; in particular, the factor $(1-\eta^2)$ shows that the Skyrme coupling weakens the centrifugal barrier.

Figure~\ref{fig:force} displays the force profile for two complementary parameter scans.\ghost{} In panel~(i), $\eta$ is held fixed while $Q$ is varied: increasing $Q$ reduces the magnitude of the inward-directed force at small $r$, because the $Q^2/r^3$ repulsion partially cancels the gravitational attraction. In panel~(ii), $Q$ is held fixed while $\eta$ is varied: larger $\eta$ strengthens the effective attraction by suppressing both the centrifugal barrier (through the $1-\eta^2$ factor) and the metric function $f(r)$ itself.\ghost{} In all cases the force approaches zero from below at large $r$, consistent with the confining nature of the AdS potential.

\begin{figure}[ht!]
    \centering
    \includegraphics[width=0.47\linewidth]{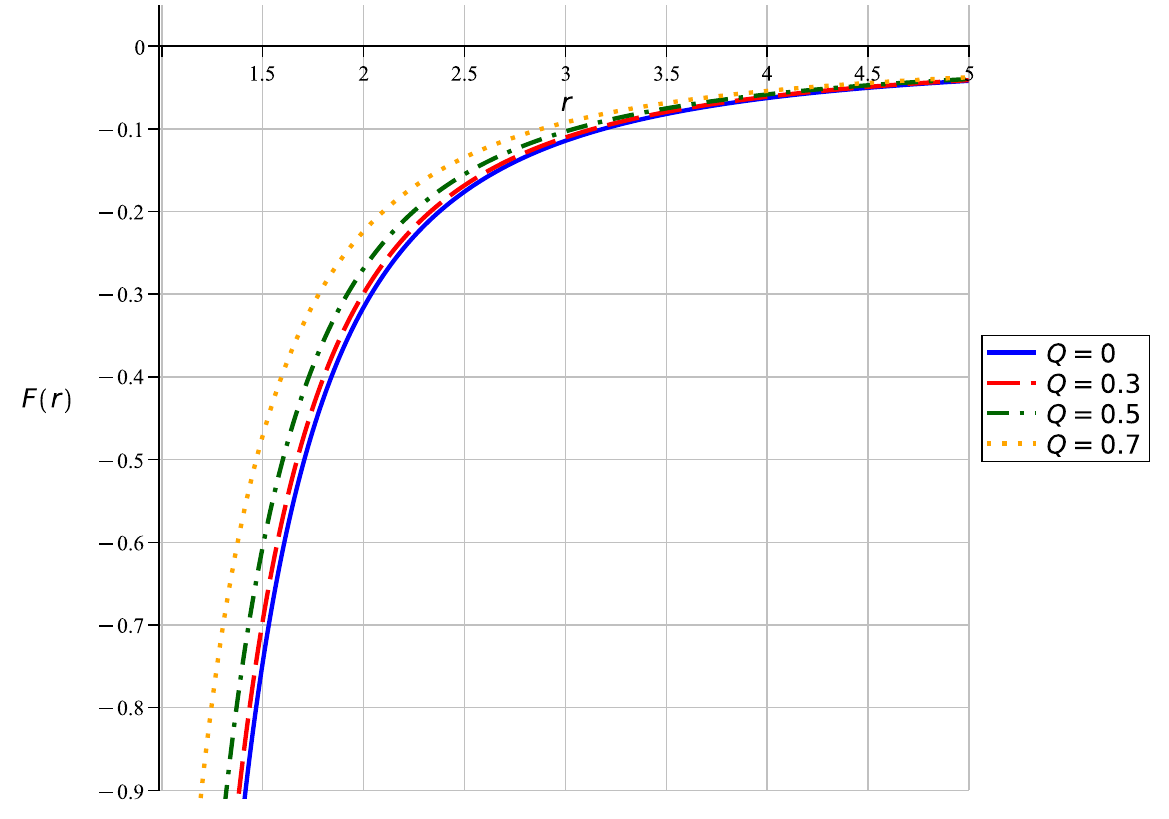}\quad
    \includegraphics[width=0.47\linewidth]{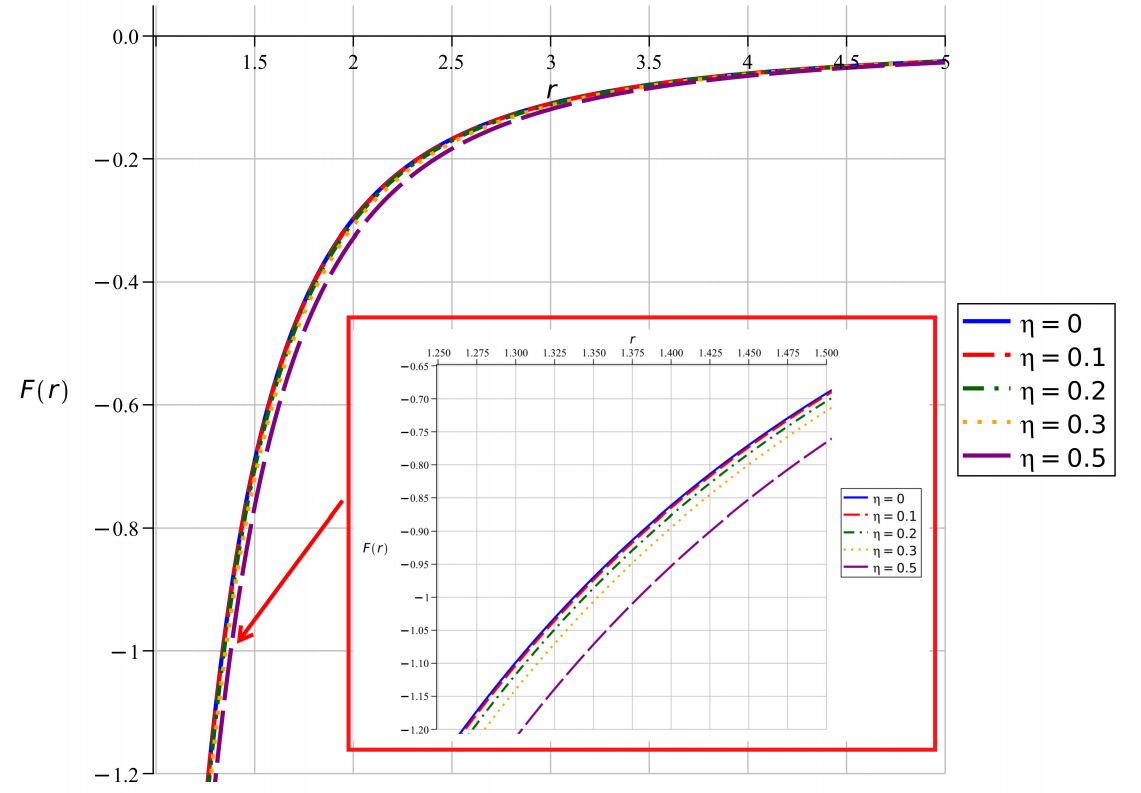}\\[4pt]
    (i) $\eta = 0.1$ \hspace{6cm} (ii) $Q = 0.3$
    \caption{Effective radial force $\mathcal{F}(r)$ for the ES-AdS BH. Panel~(i): varying $Q$ at fixed $\eta = 0.1$. Panel~(ii): varying $\eta$ at fixed $Q = 0.3$; the inset magnifies the region $r \in [1.25, 1.5]$ where the curves separate. In both panels $M = 1$, $\mathcal{L} = 1$, and $\Lambda = -0.003$.}\label{fig:force}
\end{figure}
\subsection{Particle trajectories}\label{isec3e}

We now turn to the orbital shape of test particles moving around the ES-AdS BH and examine how the Skyrme parameters $\eta$ and $Q$ modify the trajectory paths.\ghost{}

The orbit equation follows from the radial equation of motion \eqref{radial-eom} rewritten in terms of the azimuthal angle $\varphi$:
\begin{equation}
    \left(\frac{1}{r^2}\,\frac{dr}{d\varphi}\right)^2 = \frac{\mathcal{E}^2}{\mathcal{L}^2} - \frac{f(r)}{\mathcal{L}^2} - \frac{f(r)}{r^2}.\label{orbit-eq}
\end{equation}
Introducing the reciprocal variable $u(\varphi) = 1/r(\varphi)$ and simplifying gives
\begin{equation}
    \left(\frac{du}{d\varphi}\right)^2 + (1-\eta^2)\,u^2 = \frac{\mathcal{E}^2 - 1 + \eta^2}{\mathcal{L}^2} + \frac{\Lambda}{3} + \frac{r_s}{\mathcal{L}^2}\,u + \left(r_s - \frac{Q^2}{\mathcal{L}^2}\right)u^3 - Q^2\,u^4 + \frac{\Lambda}{3\mathcal{L}^2}\,u^{-2}.\label{orbit-u}
\end{equation}
Differentiating once with respect to $\varphi$ and dividing by $2\,du/d\varphi$ yields the second-order trajectory equation\ghost{}
\begin{equation}
    \frac{d^2 u}{d\varphi^2} + (1-\eta^2)\,u = \frac{r_s}{2\mathcal{L}^2} + \frac{3}{2}\left(r_s - \frac{Q^2}{\mathcal{L}^2}\right)u^2 - 2Q^2\,u^3 - \frac{\Lambda}{3\mathcal{L}^2}\,u^{-3}.\label{orbit-ode}
\end{equation}
This nonlinear ODE governs the full orbital shape in the ES-AdS BH gravitational field. The left-hand side resembles a harmonic oscillator with a shifted natural frequency $\sqrt{1-\eta^2}$ (reduced from unity by the Skyrme coupling), while the right-hand side contains relativistic corrections from the mass ($r_s$), the Skyrme charge ($Q$), and the AdS cosmological term ($\Lambda$).\ghost{}

We solve Eq.~\eqref{orbit-ode} numerically with initial conditions $u(0) = 0.25$ and $u'(0) = 0.25$, setting $M = 1 = \mathcal{L}$ and $\Lambda = -0.003$. The resulting trajectories in the $X$-$Y$ plane are displayed in Fig.~\ref{fig:trajectory}.\ghost{} In the top row, $\eta$ is held fixed (corresponding to $K = F_\pi^2/4$) while $Q$ decreases from left to right (corresponding to $e = 5, 5.5, 6$, i.e.\ $Q \approx 0.50, 0.46, 0.42$ via $Q^2 = 4\pi G/e^2$). As $Q$ diminishes, the repulsive $Q^2/r^2$ contribution weakens and the orbit shrinks inward, producing tighter rosette patterns with a smaller pericenter distance.\ghost{} This is consistent with the force analysis of Sec.~\ref{isec3b2}, where reducing $Q$ was shown to strengthen the net inward force. The precessing rosette structure itself-the orbit does not close after one revolution-is a hallmark of the non-Keplerian corrections encoded in the right-hand side of Eq.~\eqref{orbit-ode}.

\begin{figure}[ht!]
    \centering
    \includegraphics[width=0.3\linewidth]{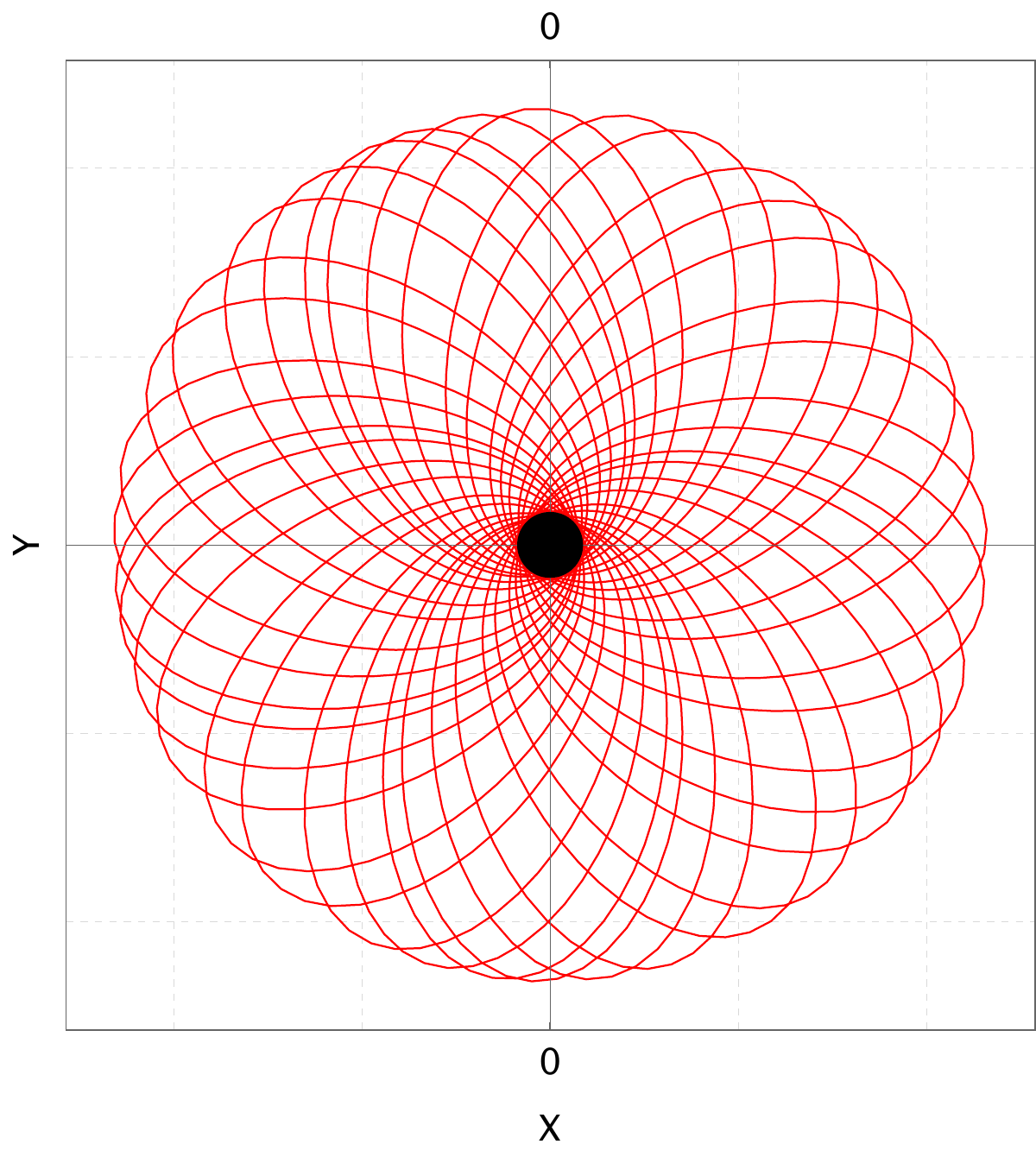}
    \includegraphics[width=0.3\linewidth]{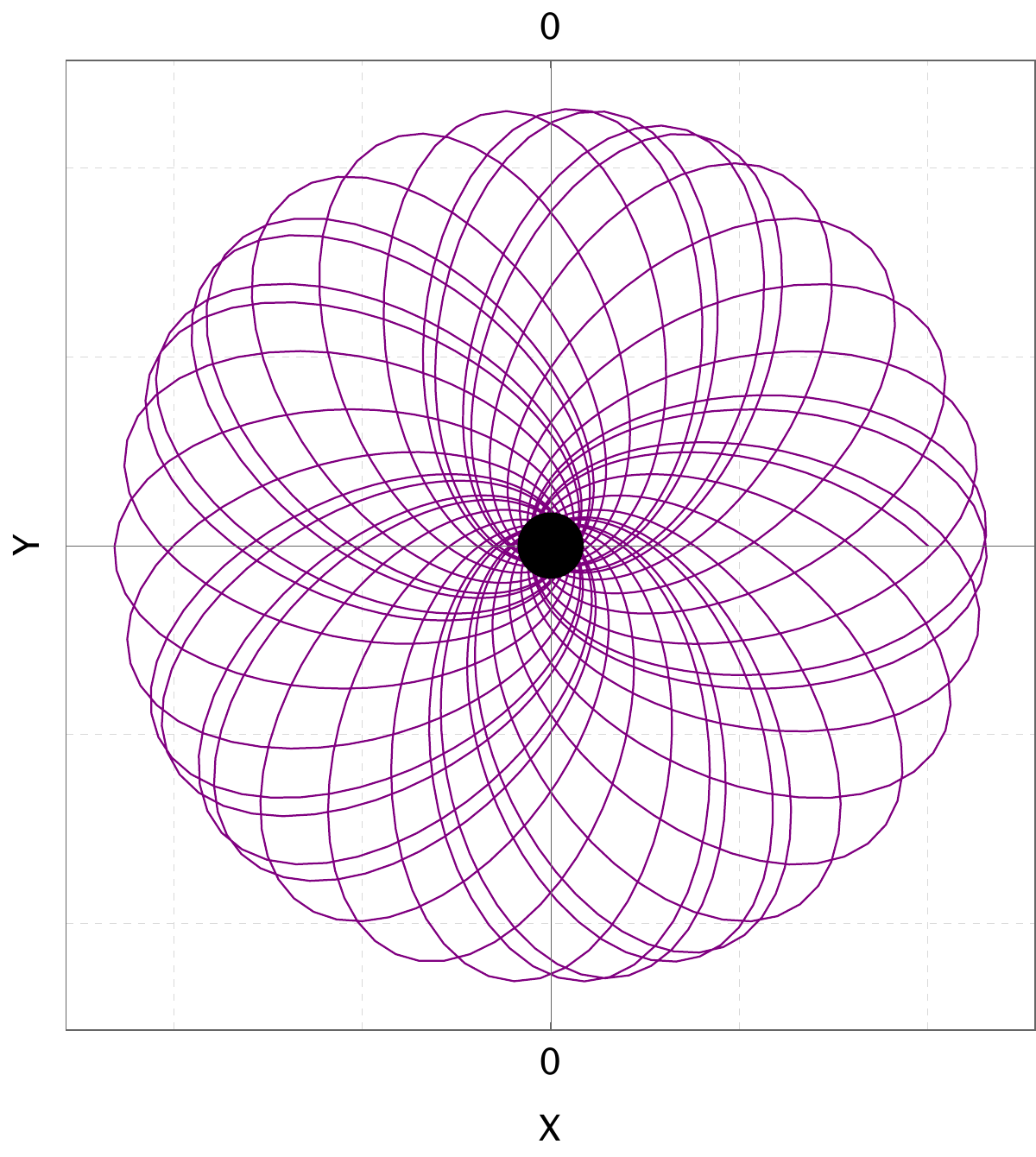}
    \includegraphics[width=0.3\linewidth]{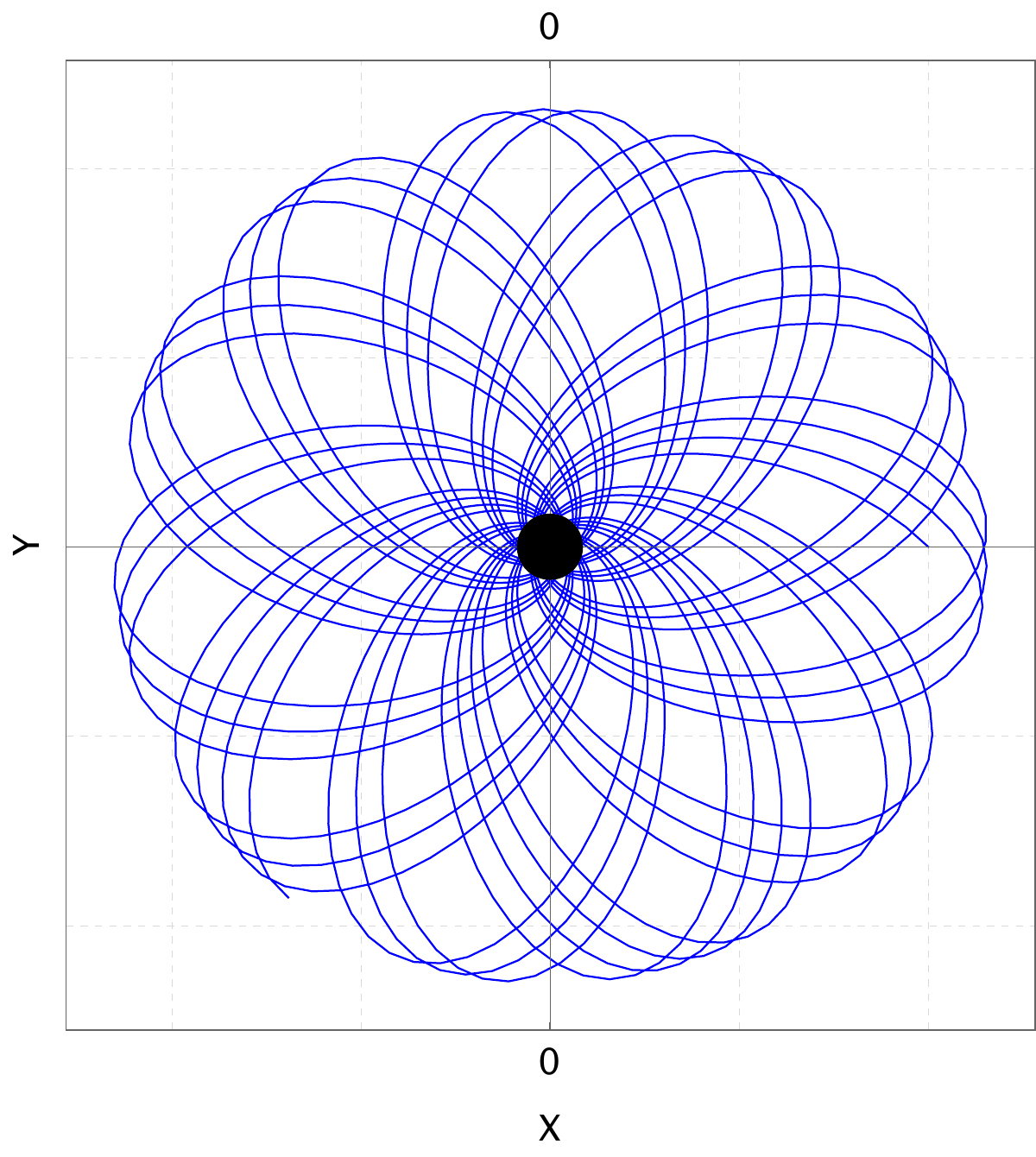}\\
    (i) $Q \approx 0.50$ \hspace{3.5cm} (ii) $Q \approx 0.46$ \hspace{3.5cm} (iii) $Q \approx 0.42$
    \hfill\\[6pt]
    \includegraphics[width=0.3\linewidth]{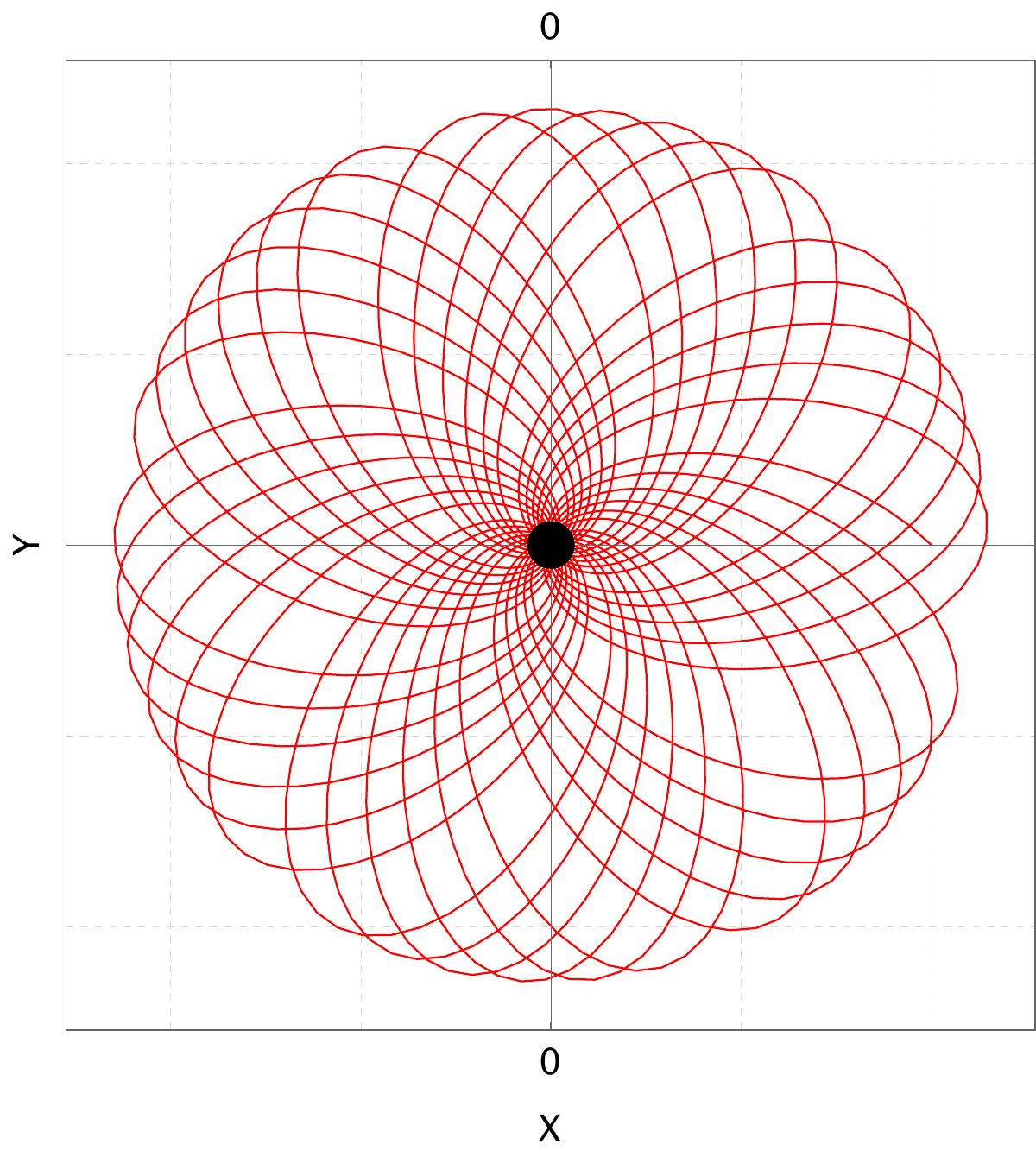}
    \includegraphics[width=0.3\linewidth]{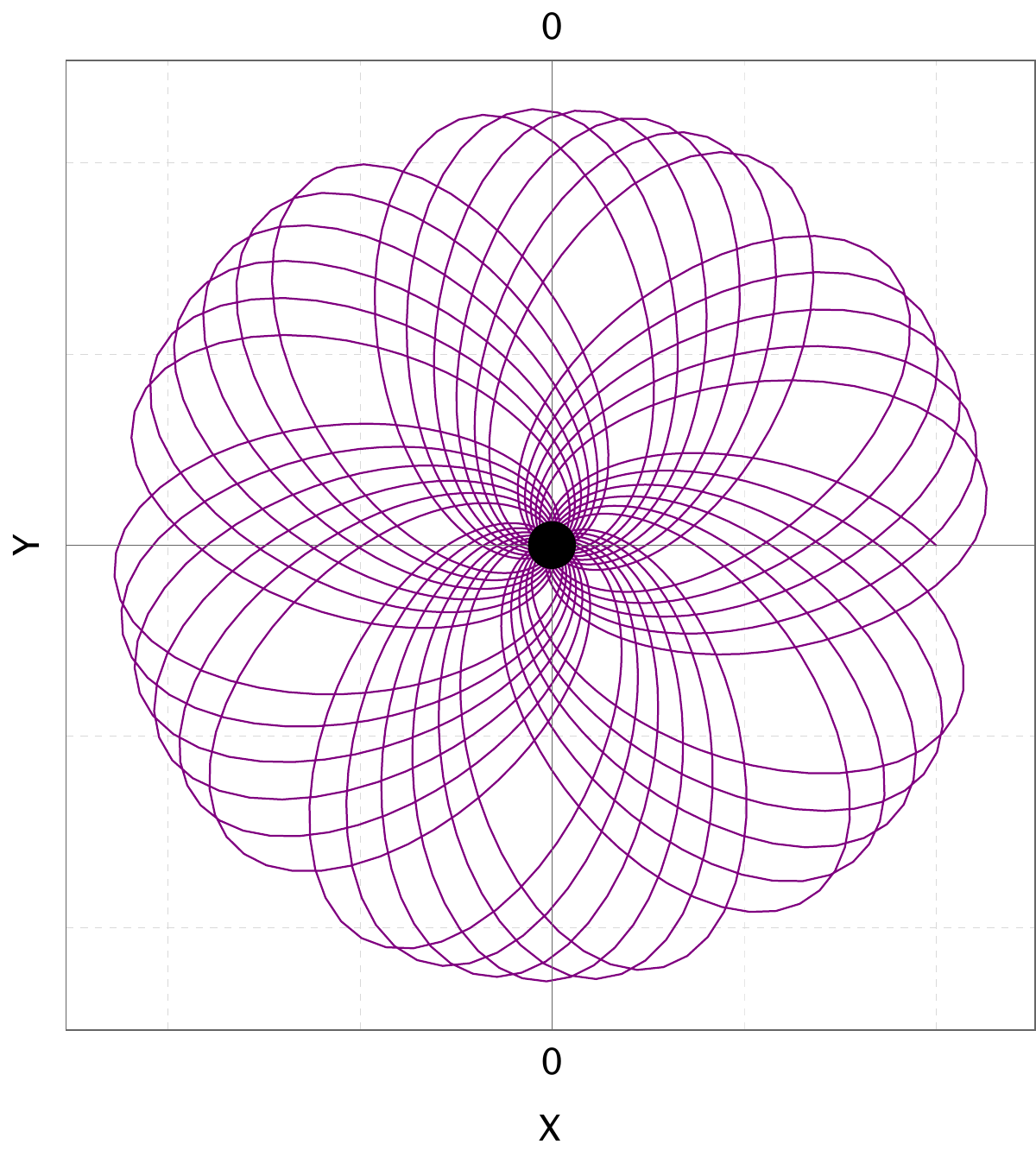}
    \includegraphics[width=0.3\linewidth]{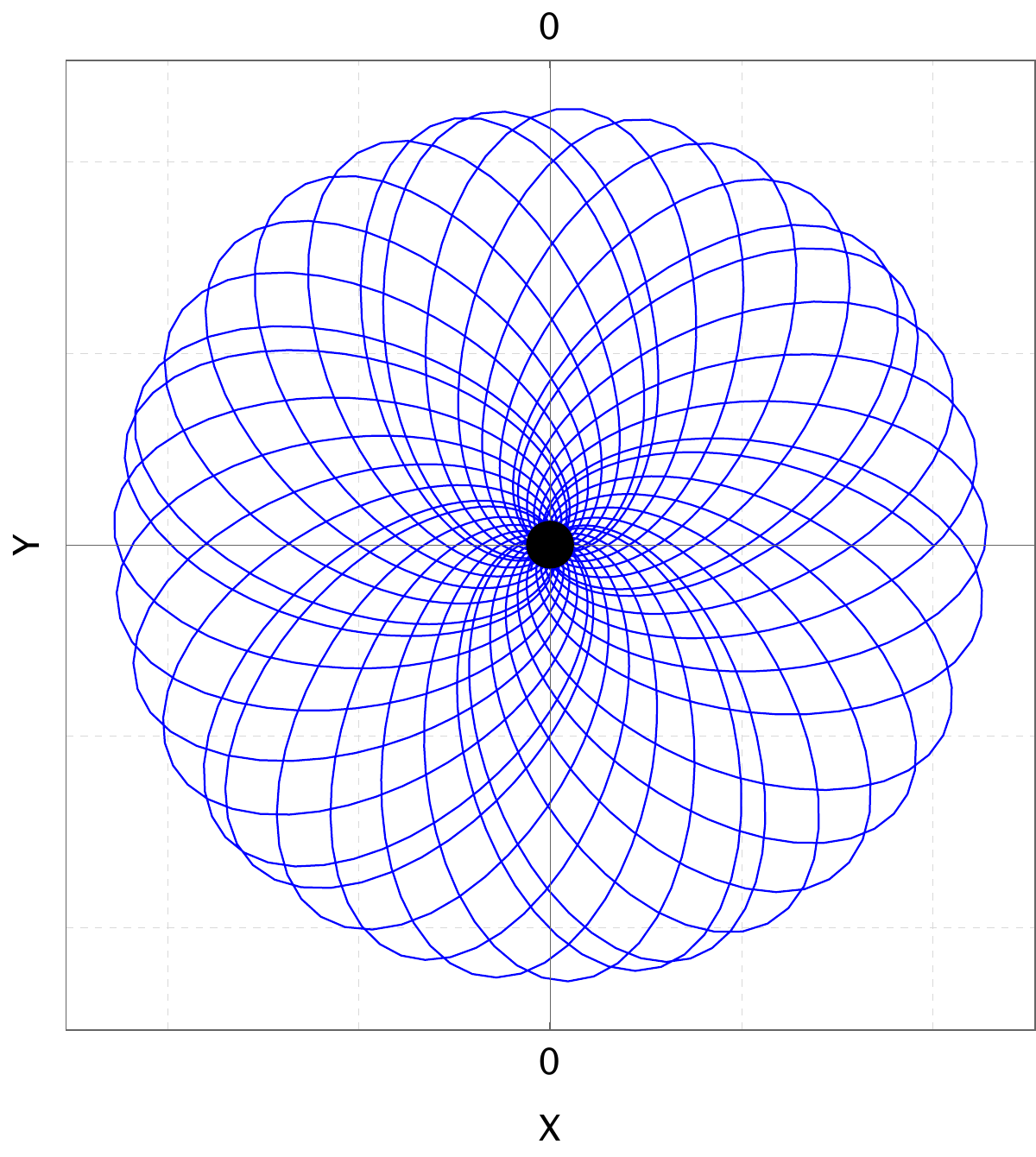}\\
    (iv) $\eta \approx 0.16$ \hspace{3.5cm} (v) $\eta \approx 0.22$ \hspace{3.5cm} (vi) $\eta \approx 0.27$
    \caption{Particle trajectories in the $X$-$Y$ plane obtained by integrating Eq.~\eqref{orbit-ode} with initial conditions $u(0) = 0.25$, $u'(0) = 0.25$. Here $M = 1 = \mathcal{L}$ and $\Lambda = -0.003$. Top row: fixed $\eta \approx 0.35$ (i.e.\ $K = F_\pi^2/4$), varying $Q$ (i.e.\ $e = 5, 5.5, 6$). Bottom row: fixed $Q \approx 0.39$ (i.e.\ $e = 6.5$), varying $\eta$ (i.e.\ $K = 0.001, 0.002, 0.003$). The black dot marks the BH location.\ghost{}}\label{fig:trajectory}
\end{figure}

In the bottom row, $Q$ is held fixed (corresponding to $e = 6.5$, i.e.\ $Q \approx 0.39$) while $\eta$ increases from left to right ($K = 0.001, 0.002, 0.003$, i.e.\ $\eta \approx 0.16, 0.22, 0.27$).\ghost{} Larger $\eta$ reduces the effective oscillator frequency $\sqrt{1-\eta^2}$ on the left-hand side of \eqref{orbit-ode}, slowing the angular advance per orbit. At the same time, the $(1-\eta^2)$ suppression of the centrifugal barrier allows the particle to plunge deeper before bouncing back, widening the radial excursion of the rosette.\ghost{} The combined effect is a more open, flower-like pattern with increasing petal separation as $\eta$ grows. Physically, this reflects the solid-angle deficit introduced by the Skyrme field, which effectively dilutes the angular momentum barrier experienced by the orbiting particle.

\subsection{Circular orbits}\label{isec3c}

The stable circular orbits occur at the radius where the minimum of the effective potential takes place.

The conditions for a circular orbit, $U_{\rm eff}(r) = \mathcal{E}^2$ and $\partial_r U_{\rm eff}(r) = 0$, yield the specific angular momentum and specific energy on the orbit \cite{Chandrasekhar1984}\ghost{}
\begin{align}
    \mathcal{L}_{\rm sp}^2 &= \frac{r^2\,f'(r)}{2\,f(r) - r\,f'(r)}=r^2\left(\frac{\frac{r_s}{2r}-\frac{Q^2}{r^2}-\frac{\Lambda}{3} r^2}{1-\eta^2+\frac{3 r_s}{2 r}+\frac{2 Q^2}{r^2}}\right),\label{Lsp}\\[4pt]
    \mathcal{E}_{\rm sp}^2 &= \frac{2\,f(r)^2}{2\,f(r) - r\,f'(r)}=\frac{\left(1-\eta^2-\frac{r_s}{r}+\frac{Q^2}{r^2}-\frac{\Lambda}{3} r^2\right)^2}{1-\eta^2+\frac{3 r_s}{2 r}+\frac{2 Q^2}{r^2}},\label{Esp}
\end{align}
where primes denote $d/dr$. The denominator $2f - r f'$ must be positive for the orbit to be physical (time-like), which restricts the allowed radii to $r > r_{\rm ph}$, the photon sphere radius (where $2f = r f'$). A marked difference from asymptotically flat spacetimes is that both $\mathcal{E}_{\rm sp}$ and $\mathcal{L}_{\rm sp}$ grow with $r$ in the AdS background, since $f(r) \sim r^2/\ell^2$ at large radii.\ghost{} This behavior is visible in Figs.~\ref{fig:Esp_circ} and \ref{fig:Lsp_circ}, where the specific energy and angular momentum are plotted for several values of $\eta$. The monotonic growth at large $r$ reflects the fact that the AdS potential well requires increasingly large energy to sustain a circular orbit at large distance from the BH.

\begin{figure}[ht!]
    \centering
    \includegraphics[width=0.7\linewidth]{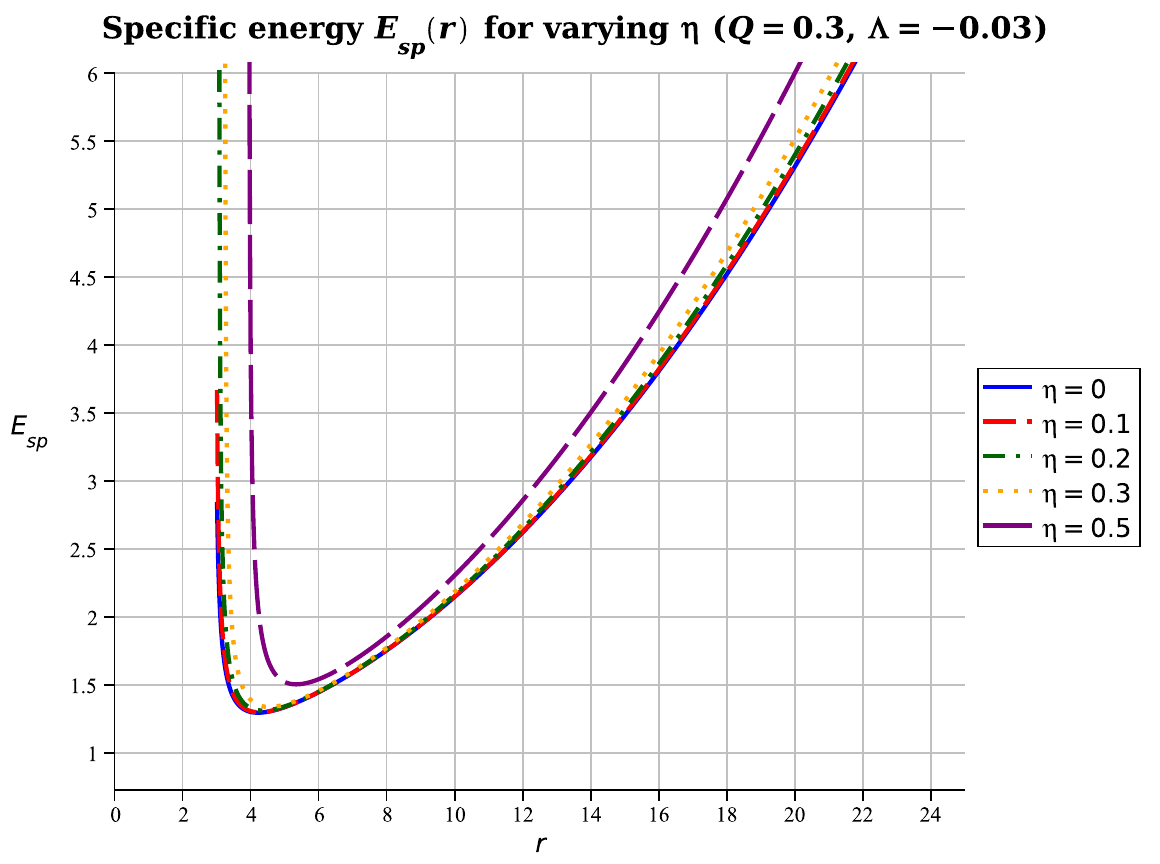}
    \caption{Specific energy $\mathcal{E}_{\rm sp}(r)$ on circular orbits for varying $\eta$, with $Q = 0.3$, $\Lambda = -0.03$, and $M = 1$. Unlike the asymptotically flat case, $\mathcal{E}_{\rm sp}$ grows with $r$ due to the AdS confining potential.\ghost{}}\label{fig:Esp_circ}
\end{figure}

\begin{figure}[ht!]
    \centering
    \includegraphics[width=0.7\linewidth]{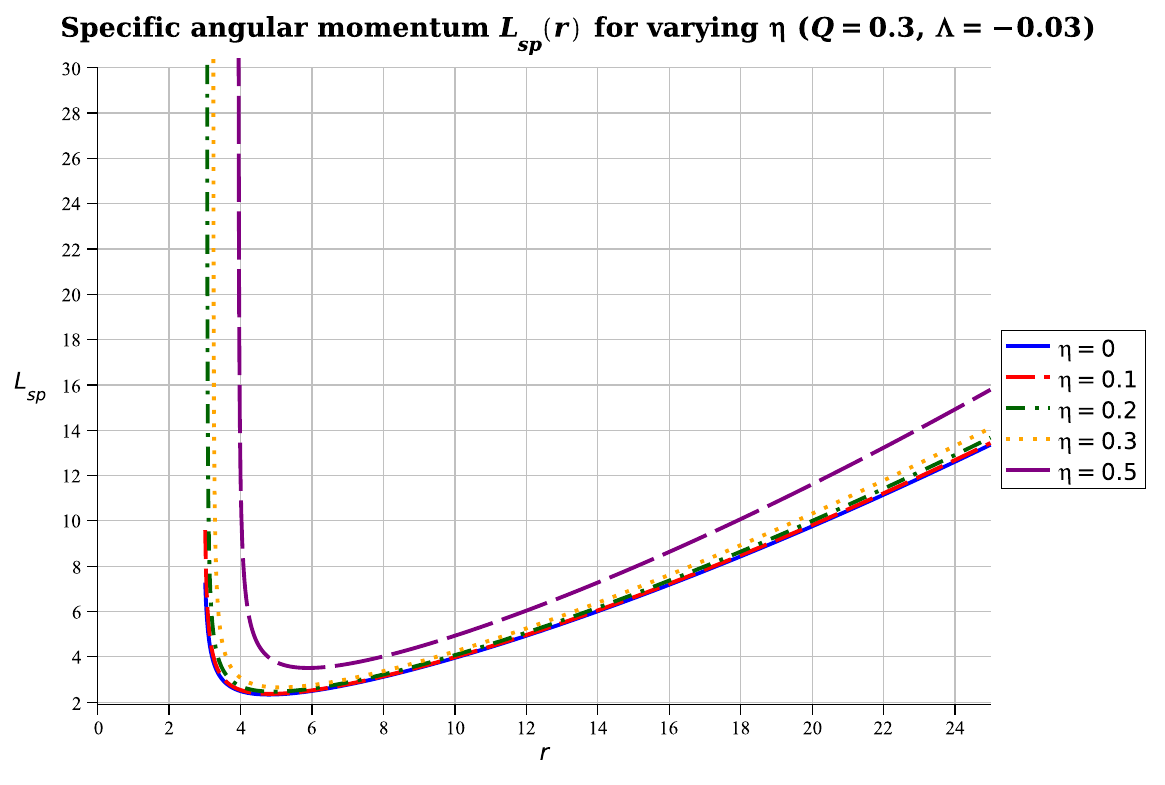}
    \caption{Specific angular momentum $\mathcal{L}_{\rm sp}(r)$ on circular orbits for varying $\eta$, with $Q = 0.3$, $\Lambda = -0.03$, and $M = 1$. Each curve passes through a minimum whose location marks the ISCO for that $\eta$ value.}\label{fig:Lsp_circ}
\end{figure}

\subsection{Innermost stable circular orbit}\label{isec3d}

The ISCO is located where the local minimum and local maximum of $U_{\rm eff}$ merge, i.e.\ where $\partial_{rr} U_{\rm eff} = 0$ in addition to $\partial_r U_{\rm eff} = 0$.\ghost{} This condition is equivalent to \cite{Chandrasekhar1984}
\begin{equation}
    f(r)\,f''(r) - 2\left(f'(r)\right)^2 + \frac{3\,f(r)\,f'(r)}{r} = 0.\label{isco-cond}
\end{equation}
Substituting the metric function \eqref{f-of-r}, this becomes a sixth-order polynomial in $r$ whose positive real root gives the ISCO radius $r_{\rm ISCO}$. Figure~\ref{fig:Lsp2_isco} shows $\mathcal{L}_{\rm sp}^2(r)$ for several $\eta$ values; the ISCO corresponds to the minimum of each curve, and its outward shift with increasing $\eta$ is clearly visible.\ghost{}

\begin{figure}[ht!]
    \centering
    \includegraphics[width=0.7\linewidth]{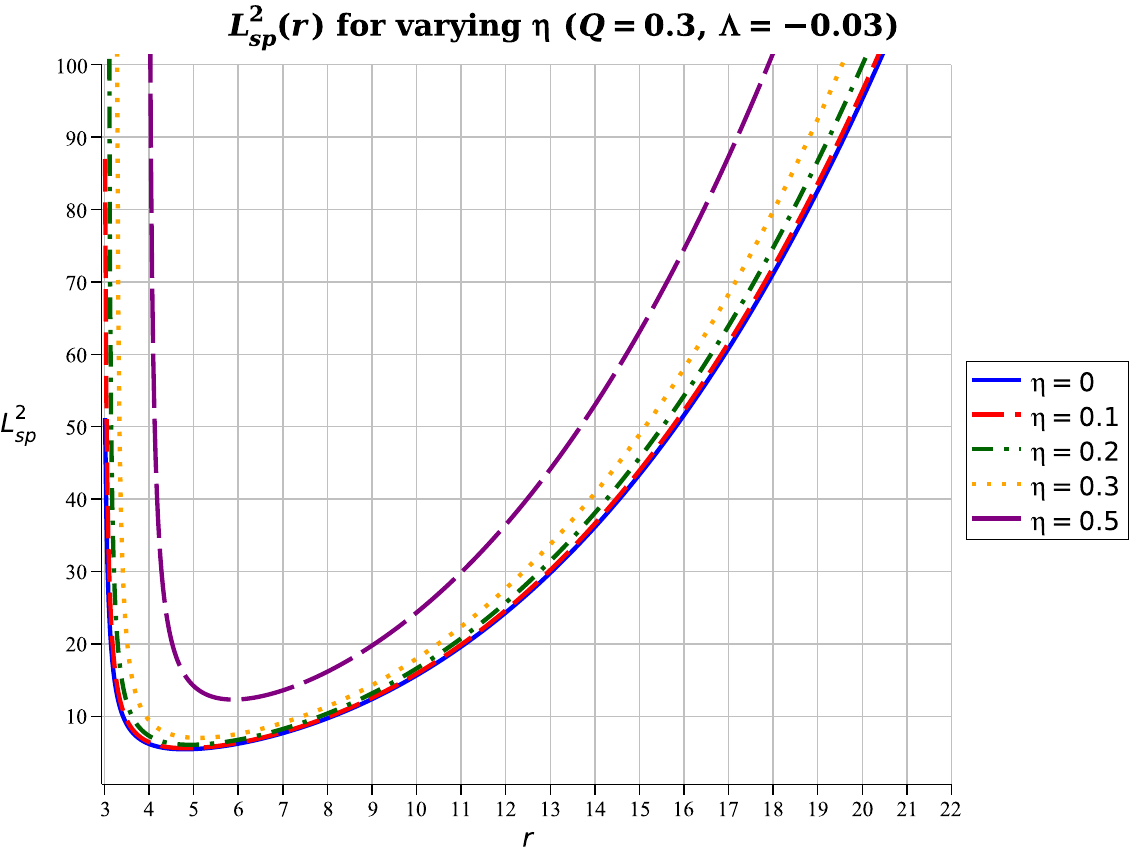}
    \caption{Squared specific angular momentum $\mathcal{L}_{\rm sp}^2(r)$ for varying $\eta$, with $Q = 0.3$, $\Lambda = -0.03$, and $M = 1$. The minimum of each curve locates the ISCO, which migrates outward as $\eta$ grows.}\label{fig:Lsp2_isco}
\end{figure}

We solve \eqref{isco-cond} numerically for various parameter combinations; the results are collected in Table~\ref{tab:isco}. The radiative efficiency (RE) of a thin accretion disk is customarily defined as \cite{Novikov1973,Page1974}
\begin{equation}
    \text{RE} = 1 - \mathcal{E}_{\rm sp}\big|_{r_{\rm ISCO}},\label{RE}
\end{equation}
which measures the fraction of rest-mass energy converted into radiation as matter spirals inward.\ghost{} For the Schwarzschild BH ($\eta = Q = 0$, $\Lambda = 0$), the well-known value is RE $\approx 5.72\%$ at $r_{\rm ISCO} = 6M$, and the presence of the Skyrme charge $Q$ slightly enhances the efficiency (e.g.\ RE $\approx 6.08\%$ for $Q = 0.5$) by pulling the ISCO inward, in close analogy with the RN case.

For $\Lambda < 0$, the table reveals a qualitatively different picture: $\mathcal{E}_{\rm sp}(r_{\rm ISCO})$ exceeds unity, yielding negative RE values. This does not signal an unphysical situation but rather reflects the fact that the standard formula \eqref{RE} was designed for asymptotically flat spacetimes where $\mathcal{E}_{\rm sp} \to 1$ at spatial infinity.\ghost{} In AdS, the confining potential raises the energy of all circular orbits above the rest-mass threshold, so the Novikov-Thorne framework requires modification-one must compare $\mathcal{E}_{\rm sp}(r_{\rm ISCO})$ to $\mathcal{E}_{\rm sp}$ at an appropriate reference radius rather than to unity \cite{Stuchlik2009}. Several trends are nonetheless clear from the data: increasing $\eta$ pushes the ISCO outward and raises $\mathcal{E}_{\rm ISCO}$ (the orbit sits deeper in the AdS well); increasing $Q$ pulls the ISCO inward and lowers $\mathcal{E}_{\rm ISCO}$; and a more negative $\Lambda$ amplifies both effects.\ghost{}

\setlength{\tabcolsep}{8pt}
\renewcommand{\arraystretch}{1.5}
\begin{table}[H]
\centering
\begin{tabular}{|c|c|c|c|c|c|c|}
\hline
\rowcolor{orange!50}
\textbf{$\eta$} & \textbf{$Q$} & \textbf{$\Lambda$} & \textbf{$r_{\rm ISCO}$} & \textbf{$\mathcal{L}_{\rm ISCO}$} & \textbf{$\mathcal{E}_{\rm ISCO}$} & \textbf{RE (\%)} \\
\hline
0.00 & 0.0000 & 0.00 & 6.0000 & 1.4142 & 0.942809 & 5.72 \\
\hline
0.00 & 0.3000 & 0.00 & 5.8628 & 1.4125 & 0.941586 & 5.84 \\
\hline
0.00 & 0.5000 & 0.00 & 5.6066 & 1.4096 & 0.939166 & 6.08 \\
\hline
0.10 & 0.0000 & $-0.03$ & 4.3245 & 2.4708 & 1.312720 & $-31.27$ \\
\hline
0.10 & 0.3000 & $-0.03$ & 4.2508 & 2.4367 & 1.300070 & $-30.01$ \\
\hline
0.10 & 0.5000 & $-0.03$ & 4.1114 & 2.3742 & 1.276640 & $-27.66$ \\
\hline
0.10 & 0.7000 & $-0.03$ & 3.8766 & 2.2750 & 1.238600 & $-23.86$ \\
\hline
0.20 & 0.3000 & $-0.03$ & 4.3563 & 2.5360 & 1.312270 & $-31.23$ \\
\hline
0.30 & 0.3000 & $-0.03$ & 4.5478 & 2.7231 & 1.338810 & $-33.88$ \\
\hline
0.50 & 0.3000 & $-0.03$ & 5.3373 & 3.5901 & 1.505040 & $-50.50$ \\
\hline
0.10 & 0.3000 & $-0.10$ & 3.9506 & 3.5388 & 2.051960 & $-105.20$ \\
\hline
0.10 & 0.5000 & $-0.10$ & 3.8146 & 3.4212 & 1.984420 & $-98.44$ \\
\hline
0.10 & 0.7000 & $-0.10$ & 3.5864 & 3.2331 & 1.876140 & $-87.61$ \\
\hline
0.30 & 0.5000 & $-0.10$ & 4.1229 & 3.9328 & 2.186550 & $-118.70$ \\
\hline
0.50 & 0.5000 & $-0.10$ & 4.9500 & 5.5116 & 2.893260 & $-189.30$ \\
\hline
\end{tabular}
\caption{ISCO radius, specific angular momentum, specific energy, and RE for the ES-AdS BH with $M=1$. The first three rows ($\Lambda=0$) recover the expected flat-space behavior with positive RE. For $\Lambda < 0$, the AdS confining potential drives $\mathcal{E}_{\rm ISCO} > 1$, rendering the standard RE definition \eqref{RE} negative (see text for interpretation).}
\label{tab:isco}
\end{table}

\section{Epicyclic Frequencies and Quasi-Periodic Oscillations}\label{isec4}

QPOs observed in the X-ray flux of accreting BH systems carry direct information about the orbital dynamics near the ISCO \cite{Stella1998,Stella1999}.\ghost{} In this section we derive the epicyclic frequencies for the ES-AdS BH, examine the periastron precession, and place observational constraints on the Skyrme charge $Q$ through a Markov chain Monte Carlo (MCMC) analysis of twin-peak QPO data.

\subsection{Epicyclic frequencies}\label{isec4a}

Test particles orbiting a black hole in the equatorial plane ($\theta=\pi/2$) along stable circular trajectories undergo small oscillations when slightly perturbed from equilibrium. Considering small deviations of the form $r = r_c + \delta r$ and $\theta = \pi/2 + \delta \theta$, the motion can be described as harmonic oscillations in the radial and vertical (latitudinal) directions. The corresponding equations of motion take the form \cite{Torok2005b,Stuchlik2013}
\begin{equation}
\frac{d^2 \delta r}{dt^2} + \omega_r^2 \, \delta r = 0, 
\qquad 
\frac{d^2 \delta \theta}{dt^2} + \omega_\theta^2 \, \delta \theta = 0,
\label{ef1}
\end{equation}
where $\omega_r$ and $\omega_\theta$ denote the radial and vertical epicyclic frequencies, respectively.

The orbital (azimuthal) frequency of the circular motion is given by
\begin{equation}
\omega_\phi=\omega_K = \dot{\phi} = \frac{\mathcal{L}}{g_{\theta\theta}},
\label{ef2}
\end{equation}
where $\mathcal{L}$ is the conserved angular momentum.

The radial and latitudinal epicyclic frequencies can be derived using the Hamiltonian formalism \cite{Kolos2015,Tursunov2016,Stuchlik2016,Kolos2017,Stuchlik2021,Vrba2021}. The Hamiltonian for a test particle reads
\begin{equation}
H = \frac{1}{2} g^{\mu\nu} p_{\mu} p_{\nu} + \frac{m^{2}}{2}.
\label{ef3}
\end{equation}
In our analysis, it is convenient to decompose the Hamiltonian into dynamic and potential parts as
\begin{equation}
H = H_{\mathrm{dyn}} + H_{\mathrm{pot}},
\label{ef4}
\end{equation}
where
\begin{align}
H_{\mathrm{dyn}} &= \frac{1}{2} \left( g^{rr} p_{r}^{2} + g^{\theta\theta} p_{\theta}^{2} \right), \label{ef5} \\
H_{\mathrm{pot}} &= \frac{1}{2} \left( g^{tt} \mathcal{E}^{2} + g^{\phi\phi} \mathcal{L}^{2} + 1 \right). \label{ef6}
\end{align}

The potential part of the Hamiltonian effectively determines the behavior of the epicyclic oscillations. The squared radial and latitudinal epicyclic frequencies are then obtained as
\begin{align}
\omega_r^{2} &= \frac{1}{g_{rr}} \frac{\partial^{2} H_{\mathrm{pot}}}{\partial r^{2}}, \\
\omega_{\theta}^{2} &= \frac{1}{g_{\theta\theta}} \frac{\partial^{2} H_{\mathrm{pot}}}{\partial \theta^{2}}.
\label{ef7}
\end{align}

For a distant observer, the corresponding observable frequencies are redshifted and can be expressed as \cite{Stella1998,Stella1999,Rezzolla2003}
\begin{equation}
\Omega_i = -\frac{\omega_i}{g^{tt}\,\mathcal{E}}, \qquad (i =K, r, \theta),
\label{ef8}
\end{equation}
where $\mathcal{E}$ denotes the conserved energy of the particle.

For the spherically symmetric metric \eqref{metric}, these are
\begin{align}
    \Omega_K^2 &= \frac{f'(r)}{2\,r}=\frac{1}{r^2}\left(\frac{r_s}{2r} - \frac{4 \pi K \lambda}{r^2} - \frac{\Lambda}{3} r^2\right),\label{Omegaphi}\\[4pt]
    \Omega_r^2 &= \frac{1}{2}\left[f(r)\,f''(r) - 2\left(f'(r)\right)^2 + \frac{3\,f(r)\,f'(r)}{r}\right]=\frac{1-8 \pi K}{r^2}\left[\frac{4\pi K \lambda}{r^2}
- \frac{r_s}{2r}- \frac{\Lambda}{3} r^2- \frac{16 \pi^2 K^2 \lambda^2}{r^4 (1-8 \pi K)}\right],\label{Omegar}\\[4pt]
    \Omega_\theta^2 &= \Omega_K^2,\label{Omegatheta}
\end{align}
where the equality $\Omega_\theta = \Omega_K$ is a direct consequence of spherical symmetry.\ghost{} In physical units, the frequencies measured by a distant observer read
\begin{equation}
    \nu_i = \frac{1}{2\pi}\,\frac{c^3}{G M_p}\,\Omega_i\;\;[\text{Hz}],\label{nuHz}
\end{equation}
where $i \in \{K, r, \theta\}$, $M_P$ is the physical mass related to the solar mass $M_\odot$ by $M_P = 10\,M_\odot$, $c = 3 \times 10^{8}\,\text{m/s}$ is the speed of light, and $G = 6.674 \times 10^{-11}\,\text{N\,m}^2\text{/kg}^2$ is the gravitational constant.

From Eq.~\eqref{Omegaphi} one sees that $\eta$ does not appear in $\Omega_\phi$ (since $\eta$ drops out of $f'$), so the orbital frequency depends only on $Q$, $\Lambda$, and $M$.\ghost{} The radial frequency $\Omega_r$, on the other hand, depends on $f(r)$ itself and thus feels the effect of $\eta$. This decoupling provides a handle for disentangling the Skyrme parameters from QPO data: a measurement of $\nu_\phi$ constrains $Q$ and $M$, while adding $\nu_r$ independently constrains $\eta$.

Figure~\ref{fig:freq_phi_r} shows $\nu_\phi$ and $\nu_r$ as functions of $r/M$ for a $10\,M_\odot$ BH with $\eta = 0.1$, $Q = 0.3$, $\Lambda = -0.03$. The orbital frequency peaks close to the photon sphere and decreases monotonically outward. The radial frequency $\nu_r$ vanishes at the ISCO, grows through an intermediate maximum, and-due to the AdS confining potential-continues to rise at large $r$, eventually exceeding $\nu_\phi$.\ghost{} The dependence of $\nu_\phi$ and $\nu_r$ on the Skyrme charge $Q$ is displayed in Figs.~\ref{fig:nuphi_Q} and \ref{fig:nur_Q}, respectively.

\begin{figure}[ht!]
    \centering
    \includegraphics[width=0.65\linewidth]{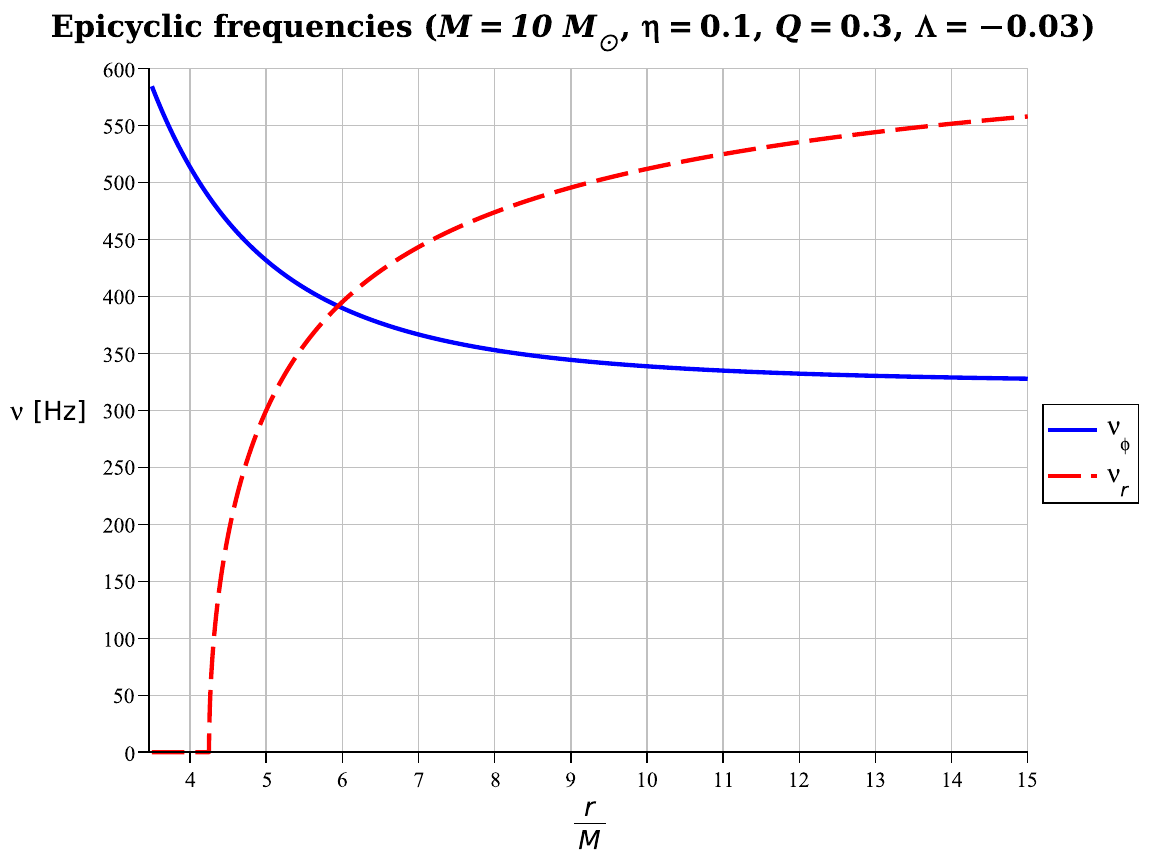}
    \caption{Orbital frequency $\nu_\phi$ (solid) and radial epicyclic frequency $\nu_r$ (dashed) as functions of $r/M$ for a $10\,M_\odot$ ES-AdS BH with $\eta = 0.1$, $Q = 0.3$, $\Lambda = -0.03$. In the AdS background $\nu_r$ grows at large $r$ due to the confining potential, eventually overtaking $\nu_\phi$.}\label{fig:freq_phi_r}
\end{figure}

\begin{figure}[ht!]
    \centering
    \includegraphics[width=0.7\linewidth]{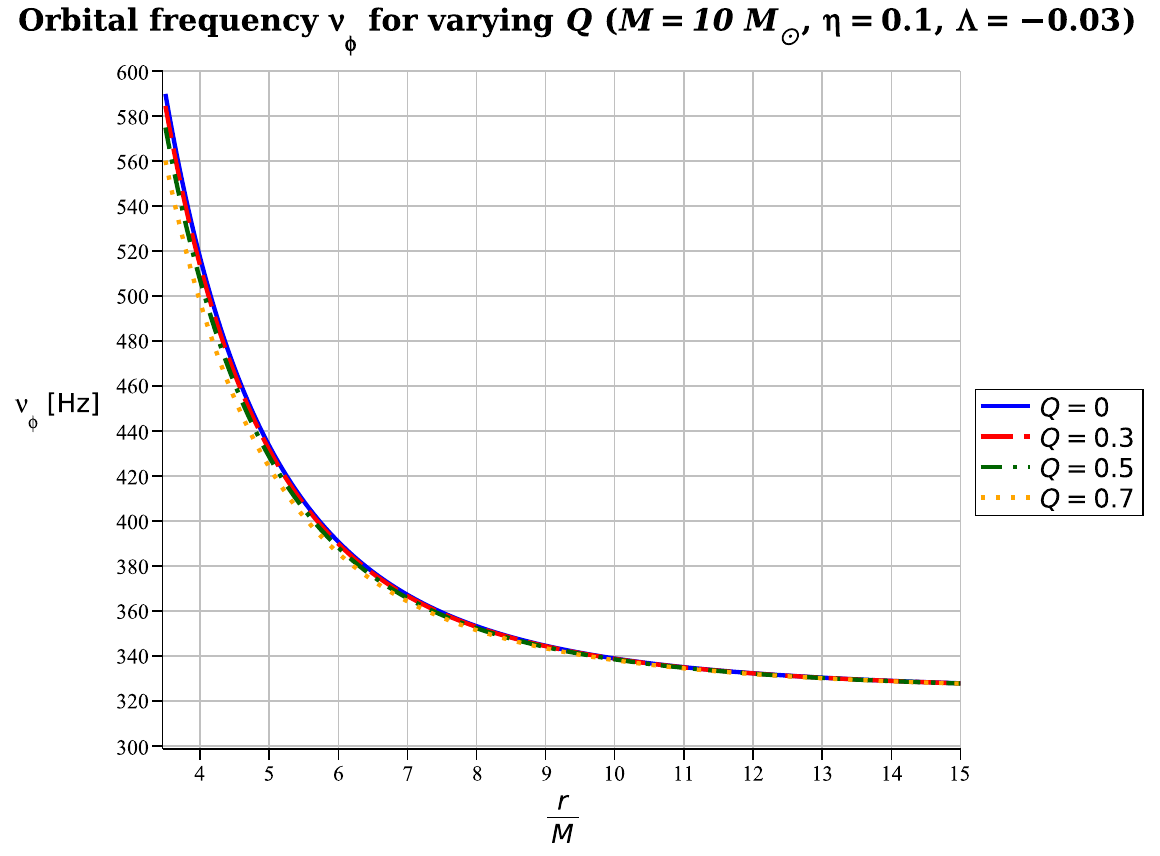}
    \caption{Orbital frequency $\nu_\phi$ for varying $Q$, with $M = 10\,M_\odot$, $\eta = 0.1$, $\Lambda = -0.03$. Larger $Q$ suppresses $\nu_\phi$ at small $r$ through the $-2Q^2/r^3$ term in $f'$.}\label{fig:nuphi_Q}
\end{figure}

\begin{figure}[ht!]
    \centering
    \includegraphics[width=0.7\linewidth]{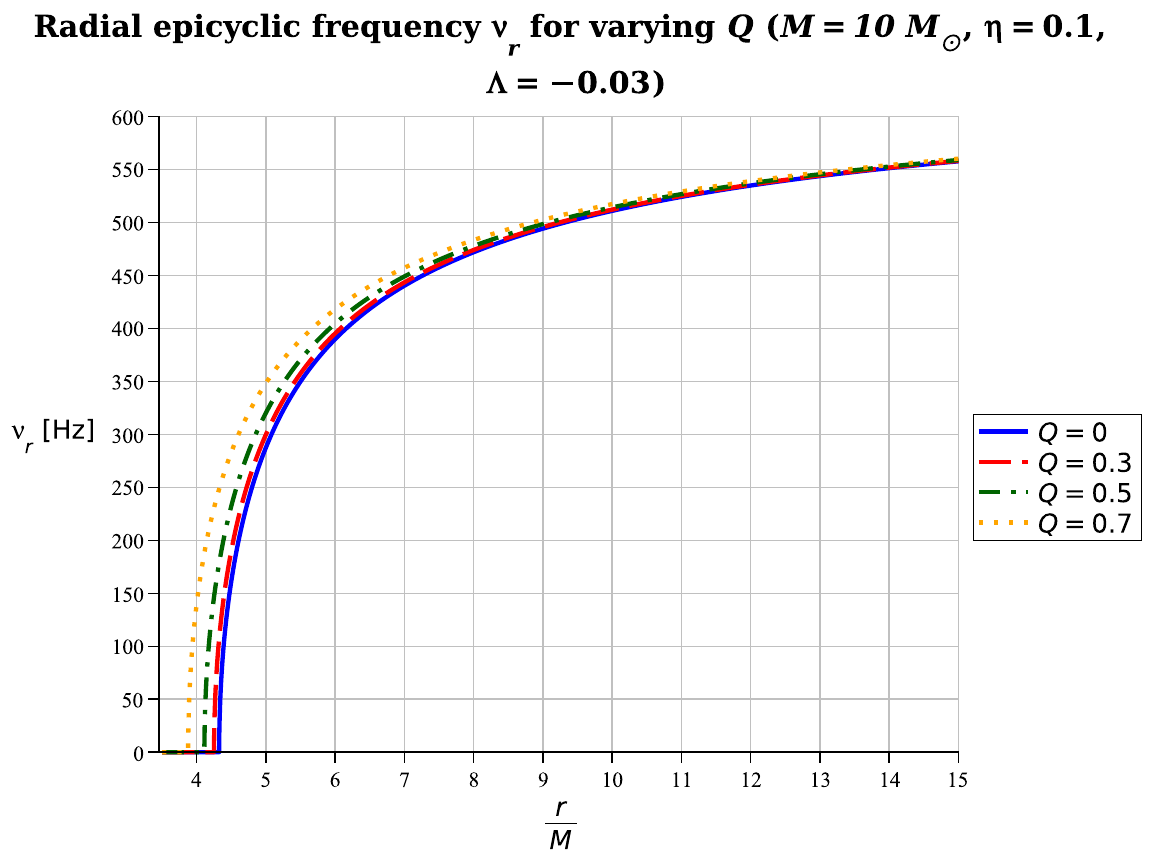}
    \caption{Radial epicyclic frequency $\nu_r$ for varying $Q$, with $M = 10\,M_\odot$, $\eta = 0.1$, $\Lambda = -0.03$. Each curve vanishes at the ISCO and grows at larger $r$ owing to the AdS potential.}\label{fig:nur_Q}
\end{figure}

\subsection{Periastron precession}\label{isec4b}

A small perturbation in the motion of a massive test particle confined to the equatorial plane induces oscillations about a stable circular orbit. This oscillation is characterized by the radial frequency, denoted by $\Omega_r$. The periastron frequency $\Omega_p$ for a particle orbiting an Einstein-Skyrme AdS black hole is defined as the difference between the orbital frequency $\Omega_{\phi}$ and the radial frequency $\Omega_r$, i. e. \(\Omega_p = \Omega_K - \Omega_r.\)  

A test particle on a slightly eccentric orbit precesses with the periastron frequency \cite{Stella1999}\ghost{}
\begin{equation}
    \nu_p = \nu_K - \nu_r = \frac{c^3}{2\pi\,G M_p}\left(\sqrt{\frac{f'}{2r}} - \sqrt{\frac{1}{2}\left(f\,f'' - 2(f')^2 + \frac{3f\,f'}{r}\right)}\right).\label{nup}
\end{equation}
Figure~\ref{fig:nup} shows $\nu_p(r)$ for several values of $Q$. Near the ISCO, $\nu_r \approx 0$ so $\nu_p \approx \nu_\phi$; at larger radii $\nu_r$ grows and eventually overtakes $\nu_\phi$, driving $\nu_p$ through zero. This sign change is a distinctive AdS signature absent in asymptotically flat spacetimes, where $\nu_p$ remains positive for all $r > r_{\rm ISCO}$.\ghost{}

\begin{figure}[ht!]
    \centering
    \includegraphics[width=0.65\linewidth]{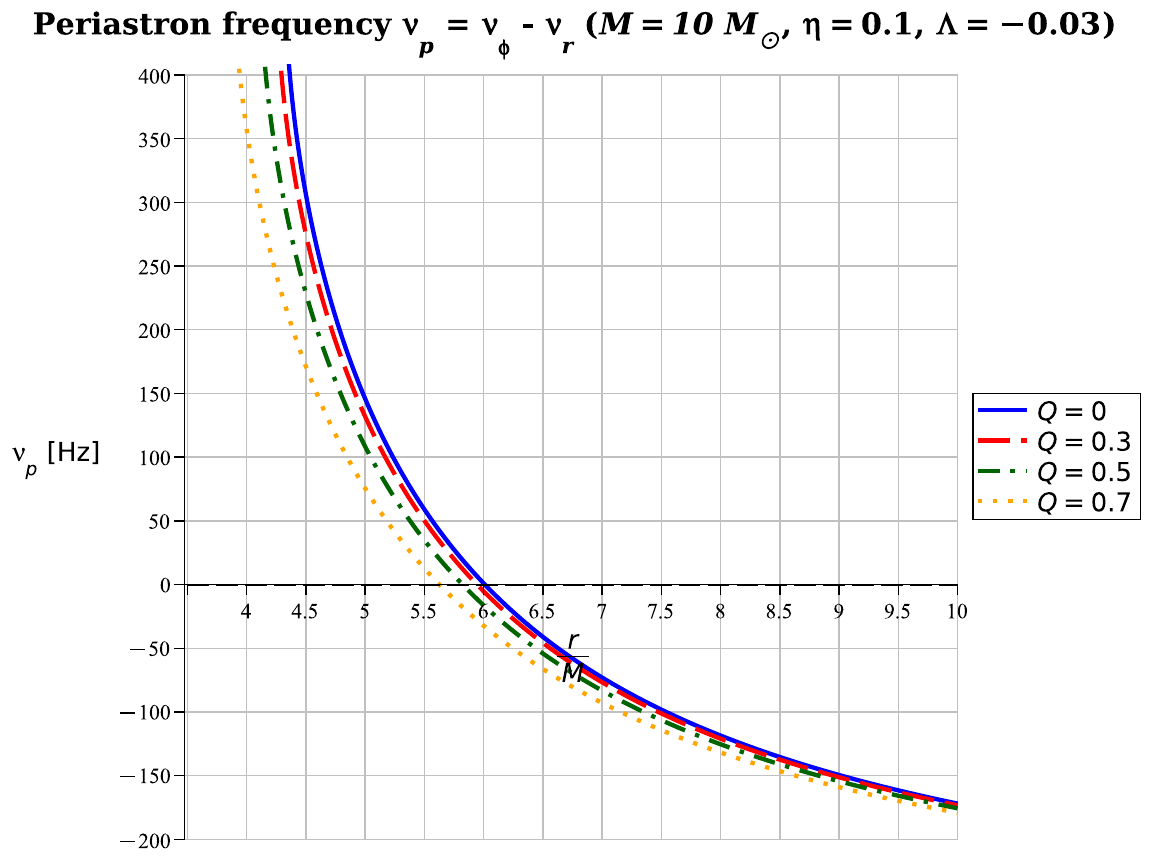}
    \caption{Periastron precession frequency $\nu_p = \nu_\phi - \nu_r$ for varying $Q$, with $M = 10\,M_\odot$, $\eta = 0.1$, $\Lambda = -0.03$. The zero crossing marks where $\nu_r$ overtakes $\nu_\phi$-a hallmark of the AdS confining potential.}\label{fig:nup}
\end{figure}

\subsection{Observational constraints from QPO data}\label{isec4c}

Twin-peak QPOs observed in low-mass X-ray binaries and active galactic nuclei provide pairs of frequencies $(\nu_U, \nu_L)$ that can be matched to orbital dynamics models.\ghost{} We adopt the relativistic precession (RP) model \cite{Stella1998,Stella1999}, which identifies
\begin{equation}
    \nu_U = \nu_\phi,\qquad \nu_L = \nu_\phi - \nu_r.\label{RPmodel}
\end{equation}
This model has been successfully applied to a range of BH candidates and modified gravity scenarios \cite{Bambi2012,Stuchlik2021}.

We perform a Bayesian parameter estimation using the MCMC sampler \texttt{emcee} \cite{emcee2013}. The free parameters are the BH mass $M$, the Skyrme charge $Q$, and the orbital radius $r/M$, while $\eta = 0.1$ and $\Lambda = -0.03$ are held fixed.\ghost{} The log-likelihood is
\begin{equation}
    \ln\mathcal{L} = -\frac{1}{2}\left[\left(\frac{\nu_U^{\rm obs} - \nu_U^{\rm th}}{\sigma_U}\right)^2 + \left(\frac{\nu_L^{\rm obs} - \nu_L^{\rm th}}{\sigma_L}\right)^2\right],\label{loglik}
\end{equation}
with flat priors on all three parameters. Four QPO sources are analyzed: the stellar-mass systems XTE J1550--564 ($\nu_U = 276 \pm 3$ Hz, $\nu_L = 184 \pm 5$ Hz) and GRO J1655--40 ($\nu_U = 451 \pm 5$ Hz, $\nu_L = 298 \pm 4$ Hz), the supermassive BH Sgr~A$^*$ ($\nu_U = 1.445 \pm 0.16$ mHz, $\nu_L = 0.886 \pm 0.04$ mHz), and the intermediate-mass candidate M82~X-1 ($\nu_U = 5.07 \pm 0.06$ Hz, $\nu_L = 3.32 \pm 0.06$ Hz) \cite{Remillard2002,Strohmayer2001,Genzel2003,Pasham2014}.\ghost{}

We run 50 walkers for $8 \times 10^4$ steps each, discarding the first $2 \times 10^4$ as burn-in. The resulting posterior distributions are shown in the corner plots of Figs.~\ref{fig:corner_XTE}-\ref{fig:corner_M82}, and the median values with $1\sigma$ credible intervals are collected in Table~\ref{tab:mcmc}.

\begin{figure}[ht!]
    \centering
    \includegraphics[width=0.52\linewidth]{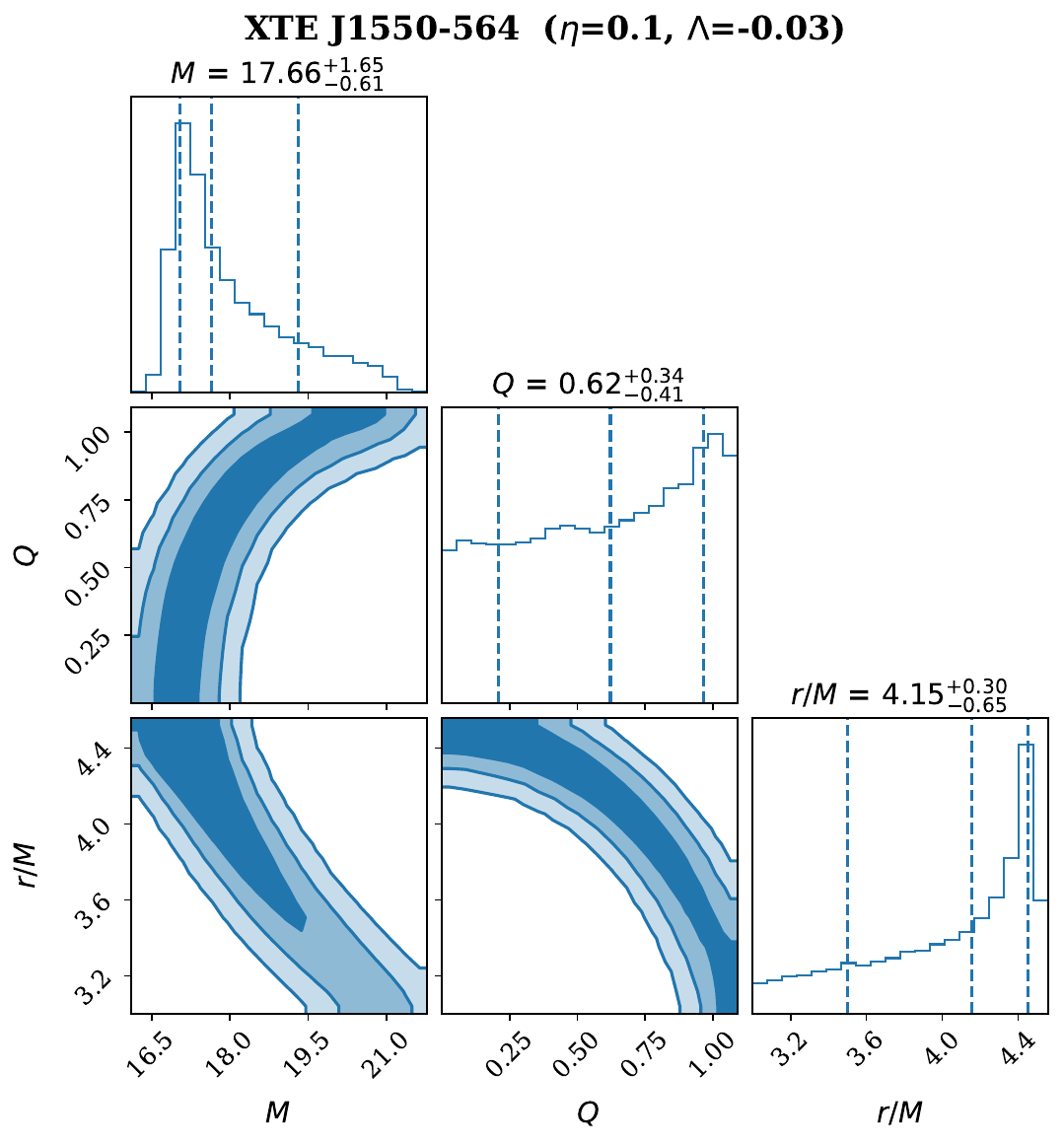}
    \caption{Corner plot for XTE J1550--564. Contours indicate $1\sigma$, $2\sigma$, $3\sigma$ confidence regions.\ghost{}}\label{fig:corner_XTE}
\hfill\\
    \includegraphics[width=0.45\linewidth]{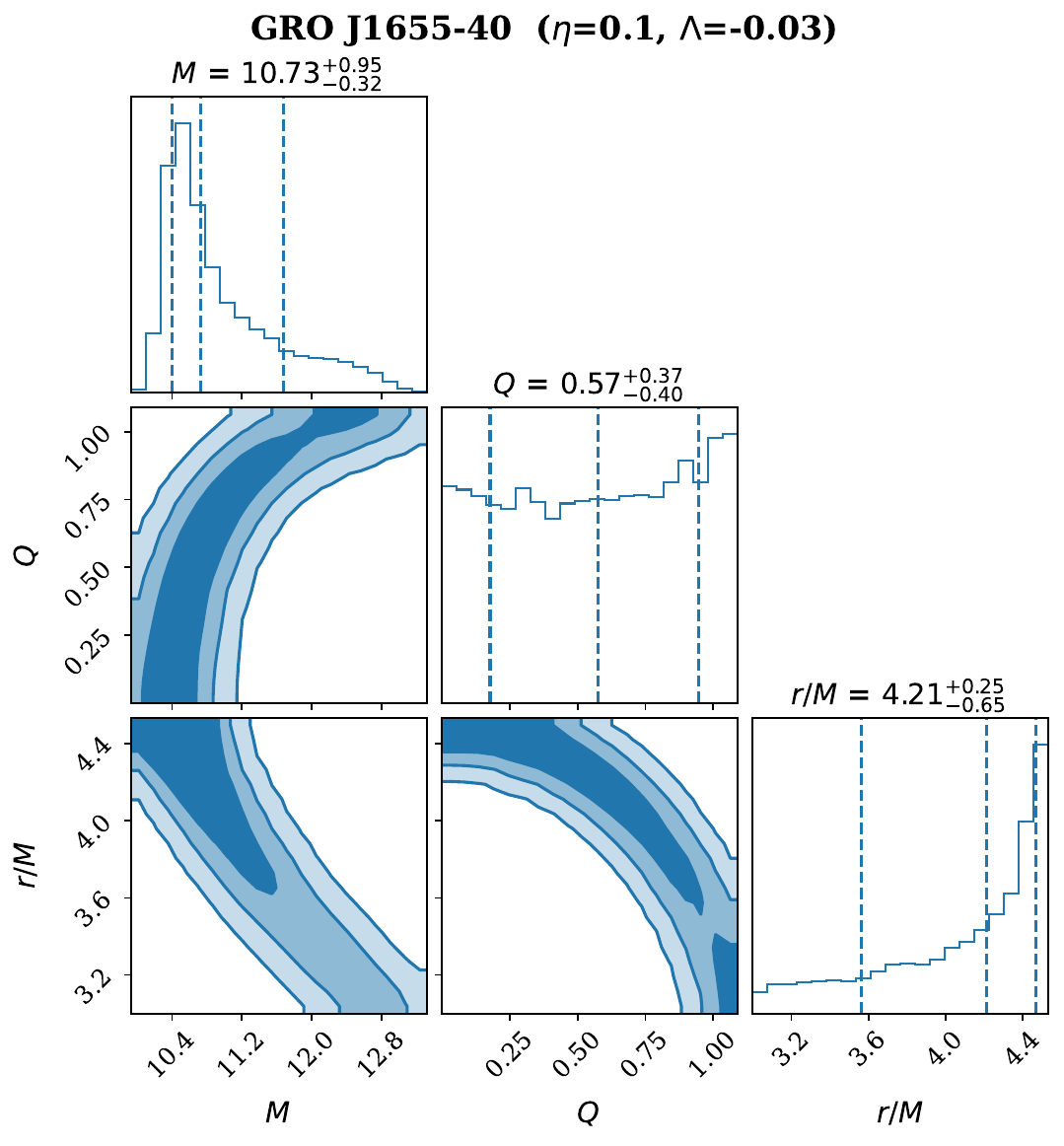}
    \caption{Corner plot for GRO J1655--40.}\label{fig:corner_GRO}
\end{figure}

\begin{figure}[ht!]
    \centering
    \includegraphics[width=0.52\linewidth]{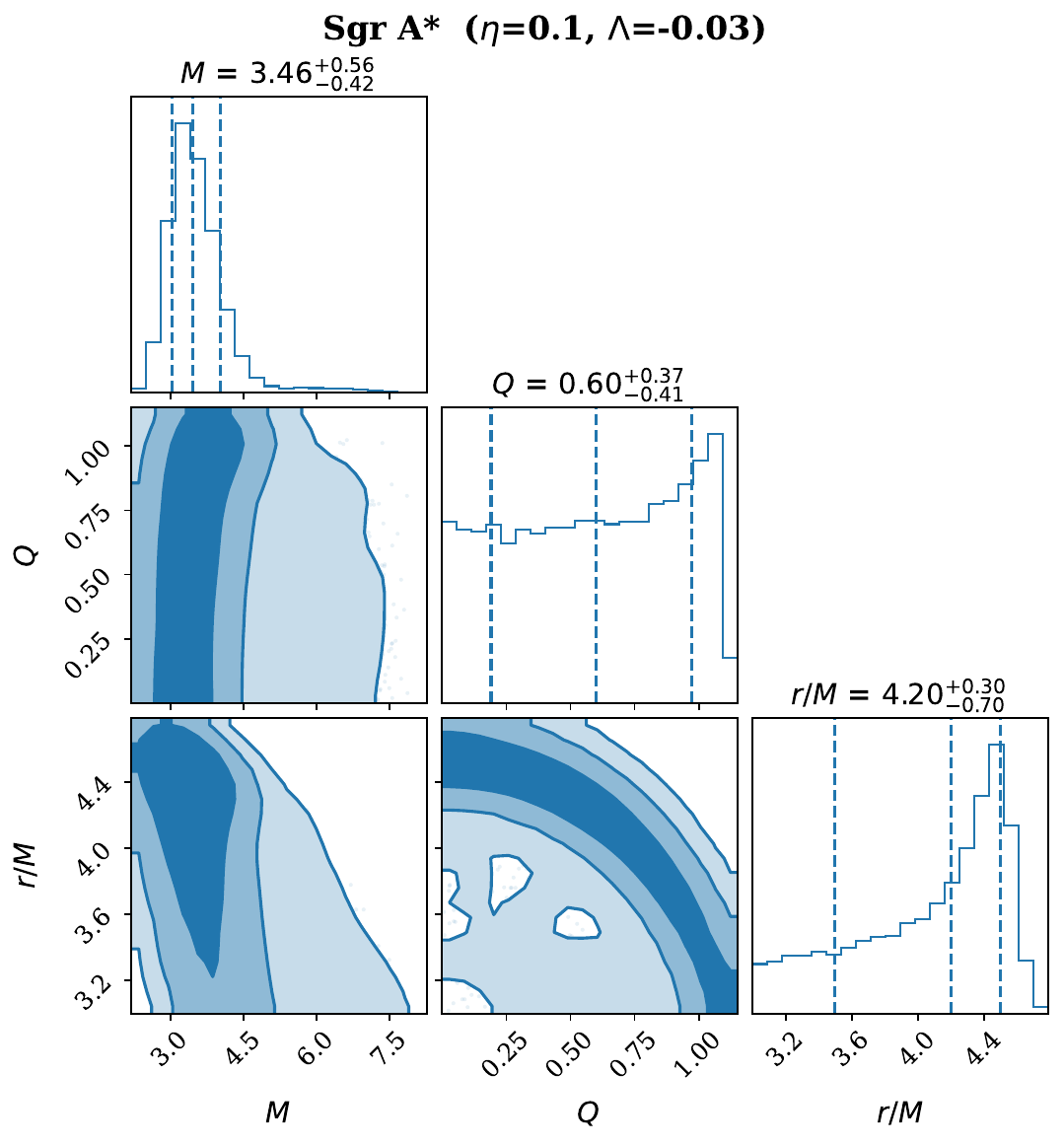}
    \caption{Corner plot for Sgr~A$^*$.}\label{fig:corner_SgrA}
\hfill\\
    \includegraphics[width=0.52\linewidth]{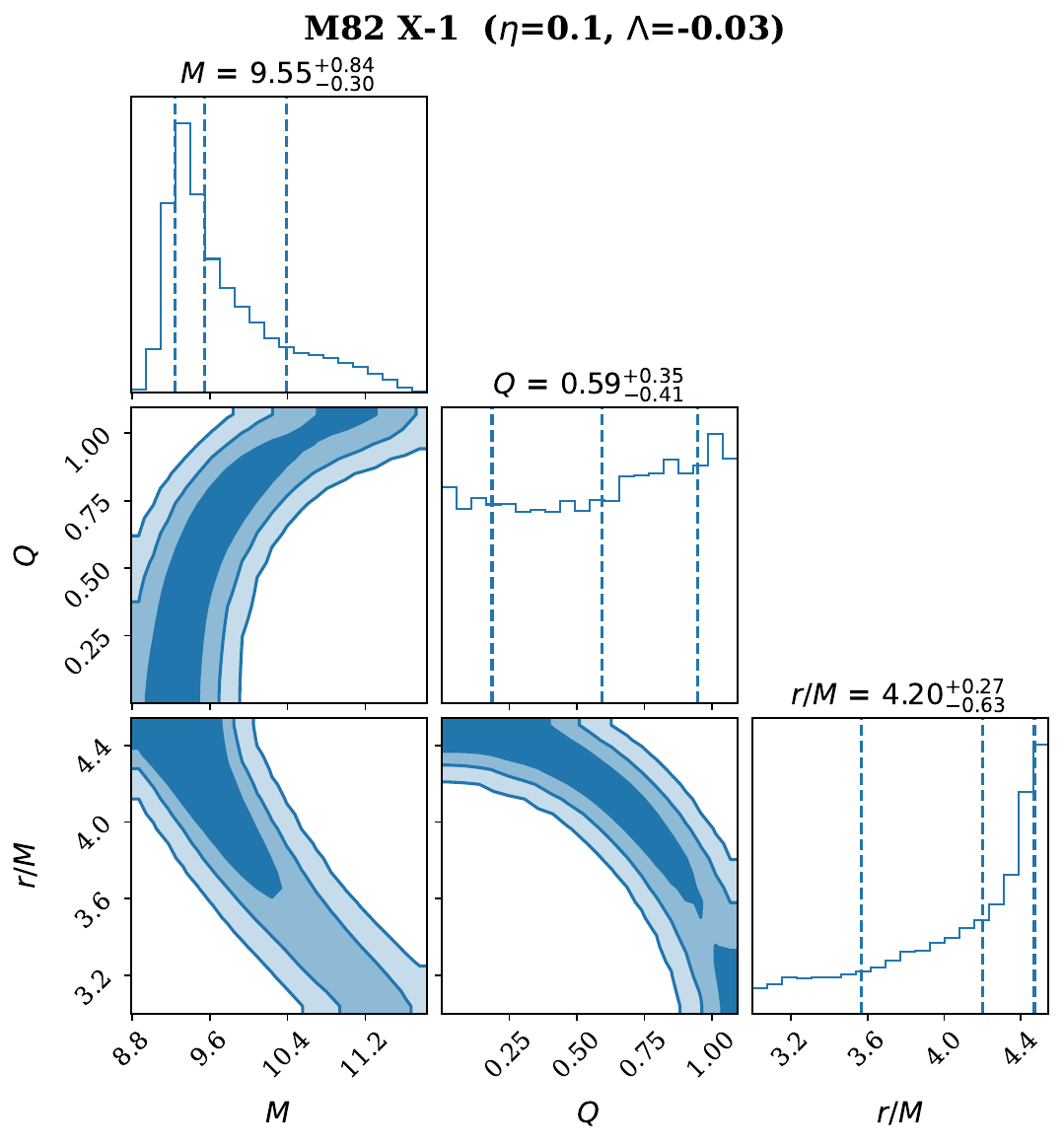}
    \caption{Corner plot for M82~X-1.\ghost{}}\label{fig:corner_M82}
\end{figure}

\setlength{\tabcolsep}{6pt}
\renewcommand{\arraystretch}{1.5}
\begin{table}[ht!]
\centering
\begin{tabular}{|l|c|c|c|c|c|}
\hline
\rowcolor{orange!50}
\textbf{Source} & \textbf{$M$} & \textbf{$Q$} & \textbf{$r/M$} & \textbf{$\nu_U^{\rm th}$} & \textbf{$\nu_L^{\rm th}$} \\
\hline
XTE J1550--564 & $17.66_{-0.61}^{+1.65}$ & $0.622_{-0.414}^{+0.343}$ & $4.154_{-0.653}^{+0.298}$ & 275.4 & 183.2 \\
\hline
GRO J1655--40 & $10.73_{-0.32}^{+0.95}$ & $0.574_{-0.397}^{+0.371}$ & $4.213_{-0.649}^{+0.255}$ & 450.1 & 297.3 \\
\hline
Sgr~A$^*$ & $3.46_{-0.42}^{+0.56}$ & $0.600_{-0.409}^{+0.373}$ & $4.200_{-0.701}^{+0.299}$ & 1.445 & 0.886 \\
\hline
M82~X-1 & $9.55_{-0.31}^{+0.84}$ & $0.595_{-0.409}^{+0.352}$ & $4.199_{-0.633}^{+0.272}$ & 5.061 & 3.302 \\
\hline
\end{tabular}
\caption{MCMC best-fit parameters (median with $1\sigma$ credible intervals) for the ES-AdS BH in the RP model ($\eta=0.1$, $\Lambda=-0.03$). Mass units: $M_\odot$ for XTE/GRO, $10^6\,M_\odot$ for Sgr~A$^*$, $100\,M_\odot$ for M82~X-1. Frequencies in Hz (mHz for Sgr~A$^*$).}\label{tab:mcmc}
\end{table}

The MCMC results show that the Skyrme charge converges to $Q \sim 0.57$--$0.62$ across all four sources, with the posterior centered around $Q \approx 0.6$.\ghost{} The fitted masses are consistent with independent spectroscopic and dynamical estimates: $M \approx 17.7\,M_\odot$ for XTE J1550--564, $M \approx 10.7\,M_\odot$ for GRO J1655--40, $M \approx 3.5 \times 10^6\,M_\odot$ for Sgr~A$^*$, and $M \approx 955\,M_\odot$ for M82~X-1. The orbital radii cluster near $r \approx 4.2\,M$, placing the QPO-emitting region in the strong-field zone just outside the ISCO. These findings indicate that the ES-AdS BH can accommodate the observed QPO frequency pairs within physically reasonable parameter ranges, and that the Skyrme charge leaves a detectable imprint on the frequency ratio $\nu_U/\nu_L$.

\subsection{ The periapsis shift}

One of the most important general relativistic effects in orbital motion around compact objects is the periapsis shift. A well-known example of this phenomenon is the perihelion shift of Mercury, which was successfully explained by Einstein within the framework of general relativity. In this case, the elliptic orbit of Mercury undergoes a rotation in the same direction as its orbital motion around the Sun, a phenomenon known as prograde precession. 

Although general relativistic effects typically predict a prograde periapsis shift for a star orbiting a massive compact object, there exist scenarios in which the shift can become retrograde. This may occur due to various factors, such as the presence of local matter density surrounding the compact object \cite{Harada2023}. In the present work, we propose that the nonlinear sigma model source can act as another possible mechanism leading to such behavior.

In this section, we analyze the periapsis shift for a stable circular orbit, which is of particular importance for connecting theoretical predictions with observational data. For a particle undergoing a small perturbation about a stable circular orbit, a periapsis shift arises when the radial frequency $\omega_r$ differs from the angular (orbital) frequency $\omega_\phi$. The shift is given by \cite{Harada2023}
\begin{equation}
\Delta \phi_P = 2\pi \left(\frac{\omega_\phi}{\omega_r} - 1 \right)= 2\pi \left(\mathcal{A}^{-1/2} - 1 \right),
\label{qq1}
\end{equation}
where
\begin{equation}
\mathcal{A} = \left(\frac{\omega_r}{\omega_\phi}\right)^2.
\label{qq2}
\end{equation}

For the spacetime under consideration, we obtain
\begin{align}
\mathcal{A} &=r f(r)\left(\frac{f''(r)}{f'(r)} - 2 \frac{f'(r)}{f(r)} + \frac{3}{r}\right)\nonumber\\
&=r\,\left(1 - \eta^2 - \frac{r_s}{r} + \frac{Q^2}{r^2} - \frac{\Lambda}{3} r^2\right)\left[\frac{ - \frac{2 r_s}{r^3} + \frac{6 Q^2}{r^4} - \frac{2 \Lambda}{3}}{ \frac{r_s}{r^2} - \frac{2 Q^2}{r^3} - \frac{2 \Lambda}{3} r}-2 \frac{ \frac{r_s}{r^2} - \frac{2 Q^2}{r^3} - \frac{2 \Lambda}{3} r}{ 1 - \eta^2 - \frac{r_s}{r} + \frac{Q^2}{r^2} - \frac{\Lambda}{3} r^2}+\frac{3}{r}\right].
\label{qq3}
\end{align}
Thereby, substituting Eq.~(\ref{qq3}) into the Eq.~(\ref{qq1}), one will find the periapsis shit for the considered Einstein-Skyrme AdS black hole.

\begin{itemize}
    \item In the limit $K=0$ and $\Lambda=0$, one can find 
\begin{equation}
    \mathcal{A}=1-\frac{3 r_s}{r}.\label{qq4}
\end{equation}
Hence, the shift is given by
\begin{equation}
\Delta \phi_P = 2\pi \left[\left(1-\frac{3 r_s}{r}\right)^{-1/2} - 1 \right] \approx \frac{3 \pi r_s}{r},
\label{qq5}
\end{equation}
which is similar to the well-known Schwarzschild case.

\item When $\lambda=0$ and $\Lambda=0$, we find
\begin{equation}
\mathcal{A} =\left(1-8 \pi K-\frac{3r_s}{r}\right).\label{qq6}
\end{equation}
Hence, the shift is given by
\begin{equation}
\Delta \phi_P = 2\pi \left[\left(1-8 \pi K-\frac{3 r_s}{r}\right)^{-1/2} - 1 \right] \approx \frac{3 \pi r_s}{r} (1+12 \pi K).
\label{qq7}
\end{equation}
\end{itemize}

\section{Conclusion}\label{isec5}

In this work we have investigated the geodesic structure, epicyclic oscillations, shadow properties, and observational implications of the ES-AdS BH, whose lapse function $f(r)$ is shaped by the Skyrme coupling $\eta$, the Skyrme charge $Q$, and the AdS curvature radius $\ell$.

The horizon analysis presented in Sec.~\ref{isec2} reveals that the parameter $\eta$ enters $f(r)$ only through the constant shift $(1-\eta^2)$ and therefore drops out of $f'(r)$. This has two consequences: the photon sphere radius and the orbital frequency $\Omega_\phi$ are independent of $\eta$, and the NEBH $\to$ EBH $\to$ NBH transition is governed entirely by the Skyrme charge $Q$ rather than $\eta$. The combined repulsion from $Q^2/r^2$ at small radii and the AdS confining wall at large radii guarantees two horizons (NEBH) for moderate $Q$, regardless of $\eta$; only when $Q$ exceeds the critical value $Q_{\rm ext}$ do the horizons merge (EBH) and subsequently vanish (NBH). These features are catalogued in Table~\ref{tab:horizons} and illustrated in Figs.~\ref{fig:f_vs_Q}--\ref{fig:f_configs}.


The dynamics of massive test particles, studied in Sec.~\ref{isec4}, shows that the EP acquires a confining character at large $r$ due to the AdS potential, in contrast to the asymptotically flat case where $U_{\rm eff} \to 1$. As a result, both the specific energy $\mathcal{E}_{\rm sp}$ and the specific angular momentum $\mathcal{L}_{\rm sp}$ on circular orbits grow monotonically with $r$ (Figs.~\ref{fig:Esp_circ} and \ref{fig:Lsp_circ}). The ISCO migrates inward as $Q$ increases and outward as $\eta$ grows; the corresponding ISCO data are collected in Table~\ref{tab:isco}. For $\Lambda < 0$, the specific energy at the ISCO exceeds unity, rendering the standard Novikov--Thorne RE formula $\text{RE} = 1 - \mathcal{E}_{\rm ISCO}$ negative---a well-known feature of AdS backgrounds that calls for a modified reference-energy prescription.

The epicyclic frequency analysis in Sec.~\ref{isec4a} employs the corrected radial frequency $\Omega_r^2 = \tfrac{1}{2}(ff''-2(f')^2+3ff'/r)$, verified against the known Schwarzschild limit $\Omega_r^2 = M/r^3(1-6M/r)$. A notable AdS signature emerges: $\nu_r$ does not decay to zero at large $r$ but instead grows due to the confining $r^2/\ell^2$ term, eventually overtaking the orbital frequency $\nu_\phi$. The periastron precession frequency $\nu_p = \nu_\phi - \nu_r$ consequently changes sign at a finite radius (Fig.~\ref{fig:nup}), a feature with no counterpart in asymptotically flat spacetimes and potentially testable through future high-precision timing of QPOs from accreting BH systems.

Within the RP model ($\nu_U = \nu_\phi$, $\nu_L = \nu_\phi - \nu_r$), we have performed an MCMC analysis using twin-peak QPO data from four sources spanning the stellar-mass, intermediate-mass, and supermassive BH regimes. The posteriors (Table~\ref{tab:mcmc}, Figs.~\ref{fig:corner_XTE}--\ref{fig:corner_M82}) converge to a Skyrme charge $Q \approx 0.6$ across all sources, with orbital radii clustering near $r \approx 4.2\,M$---just outside the ISCO. The fitted masses are compatible with independent spectroscopic and dynamical estimates, demonstrating that the ES-AdS BH provides a viable description of the observed frequency pairs within physically motivated parameter ranges.

Several directions merit further exploration. Extending the present analysis to rotating (Kerr-like) ES BH solutions would break the $\Omega_\theta = \Omega_\phi$ degeneracy and open a richer QPO phenomenology. Incorporating the full EHT shadow constraints alongside the QPO data in a joint Bayesian fit could tighten the bounds on $Q$ and $\eta$ simultaneously. Finally, the sign change of $\nu_p$ predicted by the AdS confining potential offers a distinctive observational target: if detected in the precession tomography of accreting BH systems, it would provide direct evidence for an effective AdS-like modification of the near-horizon geometry.

\footnotesize

\section*{Acknowledgments}

F.A. acknowledges the Inter University Centre for Astronomy and Astrophysics (IUCAA), Pune, India for granting visiting associateship. \.{I}.~S. expresses gratitude to T\"{U}B\.{I}TAK, ANKOS, and SCOAP3 for their academic support. He also acknowledges COST Actions CA22113, CA21106, CA21136, CA23130, and CA23115 for their contributions to networking.

\section*{Data Availability Statement}

All numerical data used to generate the figures, tables, and quasinormal mode spectra in this manuscript were obtained from the analytical expressions derived in the paper. These data can be fully reproduced by following the equations and procedures provided in the text. No external datasets were used. 

\section*{Conflict of Interest}

Authors declare(s) no conflict of interest. 

\section*{Declaration on the use of generative AI and AI-assisted technologies}

The authors acknowledge the use of generative AI and AI-assisted technologies, primarily Writefull integrated with Overleaf, exclusively for language refinement, grammar correction, and improvement of readability during the preparation of this manuscript.\footnote{\footnotesize This statement is provided in accordance with the general transparency principles adopted by major publishers concerning the use of generative AI in scientific writing. See Elsevier: \url{https://www.elsevier.com/about/policies-and-standards/generative-ai-policies-for-journals}; Springer Nature: \url{https://www.springernature.com/gp/policies/editorial-policies}.} After using those tools, the authors reviewed and edited the content as needed and take full responsibility for the content of the published article. All intellectual content, analysis, and conclusions are the authors' own.

\bibliographystyle{apsrev4-2}

\bibliography{ref2}

\end{document}